\documentclass[twocolumn]{aastex631}%, linenumbers
\pdfoutput=1 %for arXiv submission
\usepackage{amsmath,amstext}
\usepackage[T1]{fontenc}
\usepackage[figure,figure*]{hypcap}
\usepackage{mwe}
\usepackage{lipsum}
\usepackage{graphicx} 
\graphicspath{{figures/}}
\usepackage{multirow}
\usepackage{xcolor,colortbl}
\usepackage{tipa}
\usepackage{natbib}
\usepackage{hyperref}

\newcommand{\twelveCO}{$^{12}$CO }
\newcommand{\thirtCO}{$^{13}$CO }

\newcommand\HI{$\textrm{H}\scriptstyle\mathrm{I}$}
\newcommand\HII{$\textrm{H}\scriptstyle\mathrm{II}$}

\shorttitle{NGC 602: Insights from ALMA}
\shortauthors{O'Neill et al.}

\begin{document}

\title{Sequential Star Formation in the Young SMC Region NGC 602: Insights from ALMA}

\author[0000-0003-4852-6485]{Theo J. O'Neill}
\affiliation{Department of Astronomy, University of Virginia, Charlottesville, VA 22904, USA}

\author[0000-0002-4663-6827]{R\'{e}my Indebetouw}
\affiliation{Department of Astronomy, University of Virginia, Charlottesville, VA 22904, USA}
\affiliation{National Radio Astronomy Observatory, 520 Edgemont Road, Charlottesville, VA 22903, USA}

\author[0000-0002-4378-8534]{Karin~Sandstrom}
\affiliation{Center for Astrophysics and Space Sciences, Department of Physics,  University of California, San Diego, 9500 Gilman Drive, La Jolla, CA 92093, USA}

\author[0000-0002-5480-5686]{Alberto D. Bolatto}
\affiliation{Department of Astronomy, University of Maryland, College Park, MD 20742, USA}

\author{Katherine E. Jameson}
\affiliation{CSIRO Space and Astronomy, ATNF, PO Box 1130, Bentley, WA 6102, Australia}

\author{Lynn R. Carlson}
\affiliation{Experimental College, Tufts University, Medford, MA 02155}

\author[0000-0001-9338-2594]{Molly K. Finn} 
\affiliation{Department of Astronomy, University of Virginia, Charlottesville, VA 22904, USA}

\author{Margaret Meixner}
\affiliation{SOFIA Science Mission Operations/USRA, NASA Ames Research Center, Bldg. N232, M/S 232-12, P.O. Box 1, Moffett Field, CA 94035-0001}

\author[0000-0003-2954-7643]{Elena Sabbi}
\affiliation{Space Telescope Science Institute, 3700 San Martin Drive, Baltimore, MD 21218, USA}

\author{Marta Sewilo}
\affiliation{Exoplanets and Stellar Astrophysics Laboratory, NASA Goddard Space Flight Center, Greenbelt, MD 20771, USA}
\affiliation{Department of Astronomy, University of Maryland, College Park, MD 20742, USA}
\affiliation{Center for Research and Exploration in Space Science and Technology, NASA Goddard Space Flight Center, Greenbelt, MD 20771}

%%%%%%%%%%%%%%%%%%%%%%%%%%%%%%%%%%%%%%%%%%%%%%%%

\begin{abstract}
NGC 602 is a young, low-metallicity star cluster in the ``Wing'' of the Small Magellanic Cloud. We reveal the recent evolutionary past of the cluster through analysis of high-resolution ($\sim$0.4 pc) Atacama Large Millimeter/submillimeter Array observations of molecular gas in the associated $\textrm{H}\scriptstyle\mathrm{II}$ region N90. We identify 110 molecular clumps ($R <$ 0.8 pc) traced by CO emission, and study the relationship between the clumps and associated young stellar objects (YSOs) and pre-main-sequence (PMS) stars. The clumps have high virial parameters (typical $\alpha_{\rm{vir}} = $ 4 -- 11) and may retain signatures of a collision in the last $\lesssim$8 Myr between $\textrm{H}\scriptstyle\mathrm{I}$ components of the adjacent supergiant shell SMC-SGS 1. We obtain a CO-bright-to-H$_2$ gas conversion factor of $X_{CO,B} = (3.4 \pm 0.2) \times 10^{20}$ cm$^{-2}$ (K km s$^{-1}$)$^{-1}$, and correct observed clump properties for CO-dark H$_2$ gas to derive a total molecular gas mass in N90 of $16,600 \pm 2,400 \ M_\odot$. We derive a recent ($\lesssim 1$ Myr) star formation rate of $130 \pm 30 \ M_{\odot}$ Myr$^{-1}$ with an efficiency of 8 $ \pm$ 3\% assessed through comparing total YSO mass to total molecular gas mass. Very few significant radial trends exist between clump properties or PMS star ages and distance from NGC 602. We do not find evidence for a triggered star formation scenario among the youngest ($\lesssim$2 Myr) stellar generations, and instead conclude that a sequential star formation process in which NGC 602 did not directly cause recent star formation in the region is likely. 
\end{abstract}

\vspace{-3in}

%%%%%%%%%%%%%%%%%%%%%%%%%%%%%%%%%%%%%%%%%%%%%%%%%%%%%%%%%%%%%%%%%%%%%%%%%%%%%%%%%%%%%%%%%%%%%%%%%

\section{Introduction} \label{sec:intro} 

\begin{figure*}
    \centering
    \includegraphics[width=\textwidth]{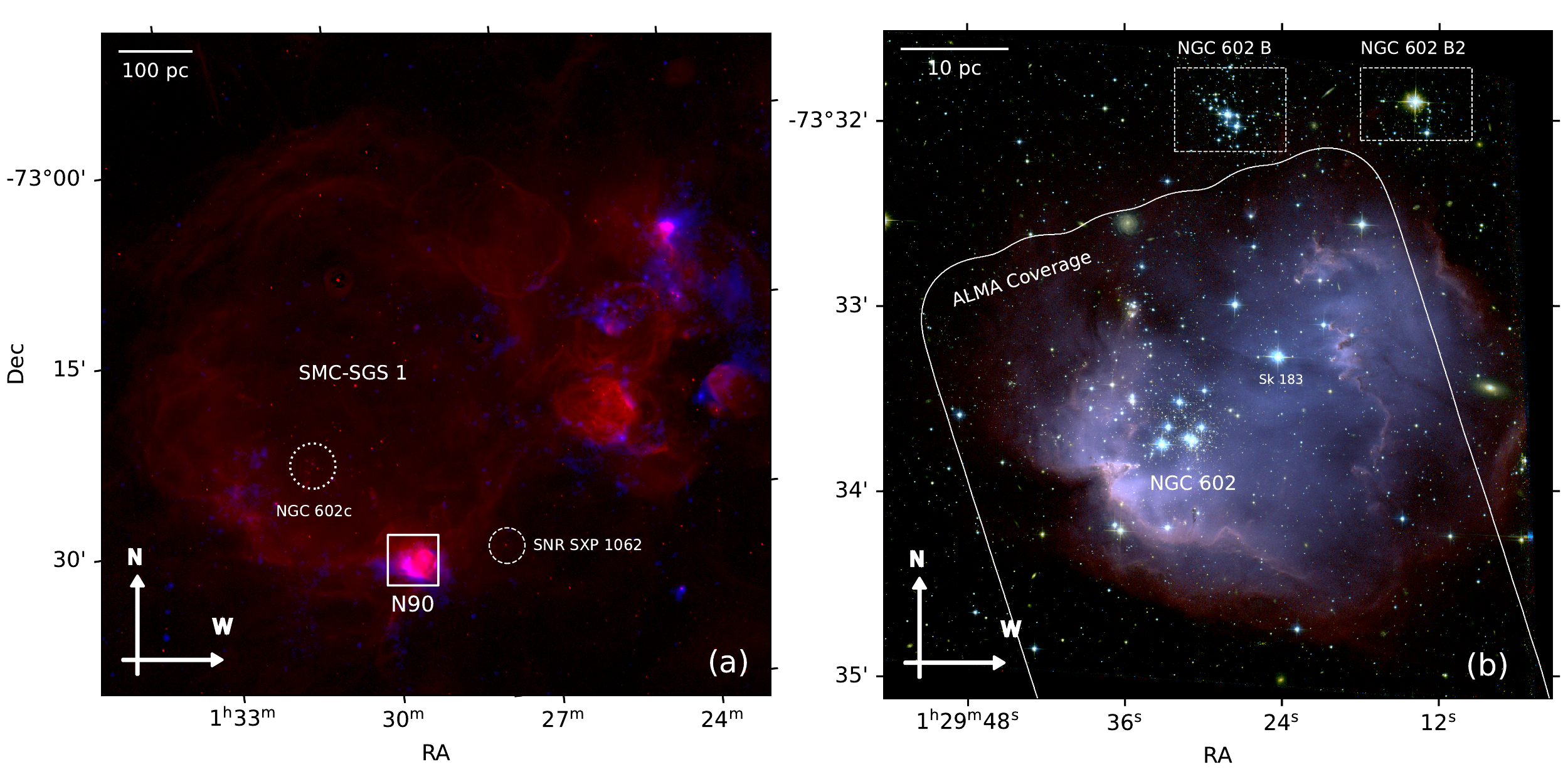}
    \caption{\textit{(a:)} Image of the supergiant shell SMC-SGS 1 in the SMC Wing.  [R, B] = [Magellanic Cloud Emission-Line Survey (MCELS) H$\alpha$ image, Herschel SPIRE 250 $\mu$m image].  The extent of the Hubble Space Telescope (HST) image of N90 shown in (b) is outlined with a white square.  SNR SXP 1062 is outlined with a white dashed circle and the cluster NGC 602c is marked with a white dotted circle.  \textit{(b:)} HST ACS image of N90.  [R, G, B] = [H$\alpha$ + F814W, F555W + F814W, F555W].  The central cluster NGC 602 is labeled, and the extent of the ALMA 12m + 7m coverage is outlined in white.  The clusters NGC 602 B and NGC 602 B2 are marked with white dashed rectangles, and the massive O3 star Sk 183 is labeled.}
    \label{fig:HubbleNGC602}
\end{figure*}

The young open cluster NGC 602 and associated \HII{} region N90 are cradled in the ``Wing’’ of the Small Magellanic Cloud (SMC) by the supergiant shell SMC-SGS 1 \citep[hereafter SGS 1,][shown in Figure \ref{fig:HubbleNGC602}a]{Meaburn1980}. 
The SMC is an ideal location to study the effects of environment on the progression of star formation: it is a low metallicity (Z $\sim 1/5$ Z$_{\odot}$, \citealt{russell_abundances_1992,lee_chemical_2005})
and low gas surface density environment (N $\sim (2-8) \times 10^{21}$ cm$^{-2}$, \citealt{leroy_spitzer_2007,welty2012}), and is located at a distance of only $\sim$60.6 kpc \citep{hilditch2005}. Under these conditions, the relationship between star-forming gas and commonly used observables like CO line emission is expected to depart from behaviors found at higher metallicities and densities.  Molecular clouds experience significant photodissociation in gas with little dust \citep{gordon_surveying_2011} and enhanced interstellar radiation fields \citep{madden_ism_2006,gordon_behavior_2008, sandstrom2010}; the fraction of H$_2$ that is “CO-dark” also increases with decreasing metallicity \citep{wolfire_dark_2010, glover_relationship_2011,szucs_how_2016}.

The Wing marks the transition between the comparatively molecule-rich inner area of the SMC and the \HI{}-dominated outer region leading to the Magellanic Bridge, which possesses even lower surface densities and metallicities than the main body of the SMC \citep{rolleston_chemical_nodate,lehner_metallicity_2008,gordon_dust--gas_2009,welty2012}.  NGC 602 and N90 have been extensively studied historically \citep[e.g.,][]{henize1956,westerlund_distribution_1964, Hodge1983,Hutchings1991}, and renewed interest in recent years has resulted in the region being remarkably well-characterized on a variety of spatial scales through a large range of wavelength regimes.  
When considered with its isolated location within the diffuse Wing, NGC 602/N90 presents a valuable opportunity for tests of star formation theory under dramatically different conditions from the Solar neighborhood.

The rate of star formation in the Wing has been increasing over the last 0.2 Gyr \citep{rubele_vmc_2015,rubele_vmc_2018}, especially in the area surrounding N90 and adjacent $\sim$500 pc diameter SGS 1 shell (Figure \ref{fig:HubbleNGC602}a).  
\citet{ramachandran_testing_2019} performed a spectroscopic investigation of OB stars within  SGS 1 and suggested that massive star formation has been ongoing in the past 100 Myr, including an extended star-formation event between 30 - 40 Myr ago.  \citet{fulmer_testing_2020} extended this work using near-UV and optical photometry and observed no radial gradient in stellar ages across SGS 1, concluding that star formation in this section of the Wing has resulted from a combination of stochastic star formation mixed with some star formation stimulated by the expansion of SGS 1. 

There is a supernova remnant SNR SXP 1062 a projected $\sim$120 pc to the west of N90 centered around a Be/X-ray pulsar binary, with age estimates ranging between (2--4) $\times \ 10^4$ years \citep{henault-brunet_discovery_2012} and (1--2.5) $\times \ 10^4$ years \citep{haberl_sxp_2012}; however, this remnant (see Figure \ref{fig:HubbleNGC602}a, diameter $\sim$25 pc) has not yet reached N90 and is unlikely to be associated with star formation in the region.  The cluster NGC 602c \citep{westerlund_distribution_1964} and associated small, very faint \HII{} region are located $\sim$190 pc to the northeast of N90 (Figure \ref{fig:HubbleNGC602}a), hosting the massive WO-type star Sk 188 and several other young, massive stars \citep{ramachandran_testing_2019}.  Given its distance, though, it too is unlikely to be directly related to recent star formation in N90.

The stellar population of N90 itself consists of a mixture of young stars concentrated around the central OB association NGC 602 and a scattered group of much older stars that are likely related to the general SMC field population. 
Figure \ref{fig:HubbleNGC602}b presents a closer view of NGC 602/N90, with two adjacent stellar concentrations to the north, NGC 602 B and NGC 602 B2, also identified; these clusters have estimated ages of up to 50--80 Myr and 47--160 Myr, respectively \citep{schmalzl_initial_2008, de_marchi_photometric_2013}. Through analysis of Hubble Space Telescope (HST) photometry, \citet{de_marchi_photometric_2013} found that one-third of pre-main-sequence (PMS) stars in N90 itself are likely $\gtrsim$30 Myr old and one-half likely younger than 5 Myr.

The cause of the formation of the central cluster NGC 602 and more recent star formation event has been a subject of debate.  \citet{cignoni_star_2009} found that the star formation rate in N90 began to increase $\sim$10 Myr ago and has peaked in the last $\sim$2.5 Myr.  Using velocity maps derived from a survey of neutral hydrogen \citep{Staveley1997} and optical and mid-IR HST data, \citet{nigra_ngc_2008} suggested that compression and turbulence from the interactions of expanding \HI{} shells $\sim$7 Myr ago is responsible for the formation of NGC 602 (with the Northern shell corresponding to SGS 1). Alternatively, \citet{fukui_formation_2020} proposed that compression resulting from a collision of two 500--600 pc radii \HI{} clouds $\sim $8 Myr ago triggered the formation of NGC 602, and that SGS 1 is the disturbed region evacuated by this cloud collision. 

Through analysis of HST optical and Spitzer Space Telescope (Spitzer) IR photometry, \citet{carlson_progressive_2007} and \citet{carlson_panchromatic_2011} concluded that NGC 602 formed $\sim$4 Myr ago, with 
a population of low-mass PMS stars forming $\sim$0.9 Myr later.  They also identified 45 candidate young stellar objects (YSOs) and proposed that star formation has propagated outwards from NGC 602 to the ``rim'' of the \HII{} region, with the youngest YSOs in N90 forming in the last $\sim$1 Myr.  
\citet{gouliermis_clustered_2007} and \citet{gouliermis_clustered_2012} analyzed the clustered spatial distribution of YSOs and PMS stars across N90 and suggested the formation of NGC 602 triggered progressive, ongoing star formation in the last 2.5 Myr in sub-clusters of PMS stars along the rim.  Alternatively, \citet{de_marchi_photometric_2013} suggested that a sequential star formation process, in which the formation of the earliest generations of young stars in the region did not significantly influence the formation of younger stellar generations, was more likely to have occurred.

%%%%%%%%%%%%%%%%%%%%%%%%%% driving questions
New, high-resolution (1.3" or 0.4 pc) Atacama Large Millimeter/submillimeter Array (ALMA) data presented here clarify the amount and nature of dense gas in N90, and the history of the region’s evolution.  In \S\ref{sec:obs}, we describe the observations and analysis methods used.  In \S\ref{sec:structure}, we analyze the structure of small molecular clumps $\sim$2--23 pc from the central cluster, as well as their association with the populations of PMS stars and YSOs.  We discuss if the clumps can reveal the formation history of the region, exploring signatures of large-scale \HI{} collisions in SGS 1 as well as evidence for feedback from NGC 602 triggering ongoing star formation along the N90 rim.  In \S\ref{sec:gasstars}, we examine common metrics of molecular cloud stability and star formation efficiency on the scales of both individual clumps and the entire region.  We conclude in \S\ref{sec:conclusions} with a discussion of the implications of our results for the evolution of N90, and compare star formation progression in N90 to solar-metallicity, higher-density environments. 

%%%%%%%%%%%%%%%%%%%%%%%%%%%%%%%%%%%%%%%%%%%%%%%%%%%%%%%%%%%%%%%%%%%%%%%%%%%%%%%%%%%%%%%%%%%%%%%%%
\section{Observations and Analysis} \label{sec:obs}

%%%%%%%%%%%%%%%%%%%%%%%%%%%%%%%%%%%%%
\subsection{ALMA Data} \label{subsec:data}

\begin{figure*}
    \centering
    \includegraphics[width=\textwidth]{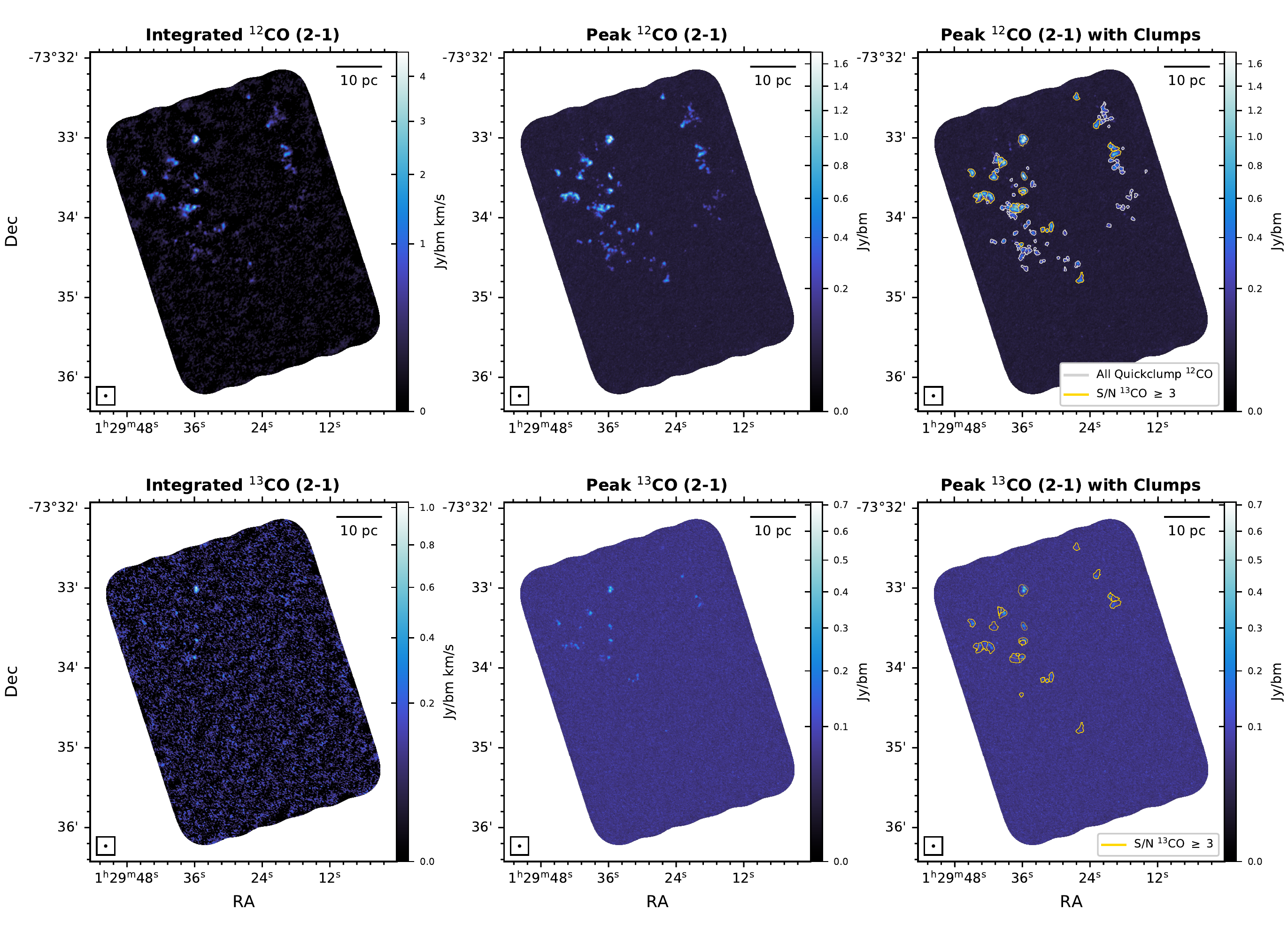}
    \caption{Images of combined 12m + 7m ALMA $^{12}$CO (2--1) and $^{13}$CO (2--1) data.  \textit{Top left:} Integrated $^{12}$CO (2--1) intensity. \textit{Top center:} Peak $^{12}$CO (2--1) intensity.  \textit{Top right:} Peak $^{12}$CO (2--1) intensity with contours of clumps identified by the {\tt{quickclump}} algorithm in grey.  Contours of the subset of these clumps with $^{13}$CO (2--1) S/N $\geq$ 3 are in gold.  \textit{Bottom left:} Integrated $^{13}$CO (2--1) intensity. \textit{Bottom center:} Peak $^{13}$CO (2--1) intensity.  \textit{Bottom right:} Peak $^{13}$CO (2--1) intensity with contours of $^{13}$CO (2--1) S/N $\geq$ 3 clumps in gold.  In all panels, the synthesized beam size is shown in the lower left corner. }
    \label{fig:Moment0s}
\end{figure*}

The NGC602/N90 region was observed by ALMA project 2016.1.00360.S.  A 150 point mosaic was observed with a 48 antenna compact configuration of the 12m array (MOUS uid://A001/X88f/X2a2), for 49 minutes on source on December 30, 2016 with a mean PWV of 1.3mm.
J0635-7516 (599 mJy at 230 GHz), J0334-4008 (432 mJy), and J0102-7546 (184 mJy) were used as bandpass, amplitude, and phase calibrator, respectively.
A slightly larger region of the sky was observed in a 60 point mosaic using the 7m ACA eight times between October 12 and 24, 2016, for a total of 416 minutes on source, mostly at a PWV of $\sim$ 0.5 mm.  J0006-0623 (3.6 Jy at 230 GHz) or J0522-3627 (2.8 Jy) were used as bandpass calibrator, Uranus as amplitude calibrator, and J0450-8101 (1.6 Jy) as phase calibrator.  
The spectral setup contains 3 spectral windows with 122.07 kHz channels each, centered on the $^{12}$CO, $^{13}$CO, and $^{12}$C$^{18}$O J=2--1 transitions.  These windows have 1920, 2048 channels and 234.4 MHz, 250 MHz for the 12m and 7m arrays, respectively.  Additionally observed was a 2 GHz wide, 128 channel spectral window centered at 232.86 GHz.  The native spectral channel spacing is 0.17 km/s, with a resolution for $^{12}$CO of 0.184 km/s.  

The data were processed with the ALMA Pipeline-CASA56-P1-B v42866 released with CASA 5.6.1-8 \citep{McMullin2007_casa}, using the default recipes.  Standard flagging resulted in 27\% of the 12m data and 35\% of the 7m data being flagged. The imaging stages of the ALMA pipeline correctly detected strong line emission and subtracted from the visibilities a linear fit to the continuum, excluding the line spectral ranges. 

Calibrated visibilities from both arrays were imaged together using CASA::tclean,
Total Power data were also observed with ALMA, and processed with Pipeline-Cycle4-R2-B packaged with CASA 4.7.0. The Total Power data were feathered with the interferometer data, but the interferometers alone recovered $>$95$\pm$5\% of the $^{12}$CO 2--1 flux and 100$\pm$10\% of the $^{13}$CO 2--1 flux in the Total Power image  (i.e., within the absolute calibration uncertainty), and the feathered image has higher noise on large angular scales, so we use the interferometer-only images for the clump analysis presented here.  The combined 12m + 7m integrated intensity images in $^{12}$CO 2--1 and $^{13}$CO 2--1 are shown in Figure \ref{fig:Moment0s}.

%%%%%%%%%%%%%%%%%%%%%%%%%%%%%%%%%%%%%
\subsubsection{Clump Extraction} \label{subsec:extract}

Molecular cloud emission is often decomposed into discrete ``clumps'' (R $\sim 1$ pc) or ``cores'' (R $\lesssim 0.1$ pc) to enable analysis of the characteristics of the complex structures within the clouds.  
Non-hierarchical clump identification methods like {\tt clumpfind} \citep{williams_determining_1994} or {\tt quickclump} \citep{Sidorin2017_quickclump} segment position-position-velocity (PPV) cubes by identifying local maxima and assigning adjacent pixels above a minimum intensity $I_{min}$ with a minimum intensity difference $\delta I$ between the local maximum and the highest adjacent saddle point to discrete clumps.  
Diffuse molecular gas is typically not visible in observable CO emission in the SMC, as the low metallicity, low dust-to-gas ratio, and higher interstellar radiation fields of the SMC result in observed CO emission being segmented into more discrete, ``clumpier'' structures \citep[e.g.,][]{muraoka_alma_2017, jameson_first_2018} than in the Milky Way or Large Magellanic Cloud (LMC) where filamentary structures are more common \citep[e.g.,][]{saigo_kinematic_2017, indebetouw_structural_2020}. Non-hierarchical methods like {\tt clumpfind} are then more suited to environments like the SMC due to this tendency towards discrete CO structures.  

We used a version of the {\tt quickclump} python implementation\footnote{\url{https://github.com/vojtech-sidorin/quickclump/}} modified to include a parameter defining a required minimum peak intensity of a clump, $I_{minpk}$ \citep{indebetouw_structural_2020}\footnote{\url{https://github.com/indebetouw/quickclump}}.  The addition of this parameter ensures that clumps with relatively high signal to noise ratio (S/N) peaks are able to have emission in their envelopes assigned down to the level of the noise, which allows for the emission of the clump to be captured more completely, while avoiding introducing many additional clumps with low S/N peaks.  

We applied the modified {\tt quickclump} algorithm to the $^{12}$CO observations, with a minimum intensity $I_{min}$ = 4$\sigma$, minimum change in intensity between leaves $\delta I = 3 \sigma$, minimum peak intensity $I_{min,pk}$ = 8$\sigma$, and minimum number of pixels $n_{min,pix}\simeq 2$ beams.  Finally, we required that each clump have $^{12}$CO velocity dispersions (as calculated in \S\ref{S:clump_props}) greater than the spectral resolution for that line ($\sigma_v \gtrsim$ 0.18 km s$^{-1}$).  This yielded a total of 110 clumps.

%%%%%%%%%%%%%%%%%%%%%%%%%%%%%%%%%%%%%
\subsubsection{Molecular Column Density} \label{subsec:coldensity}

We assumed local thermal equilibrium (LTE) conditions to calculate column densities for clumps with significant $^{13}$CO emission.  Since many of the clumps identified in $^{12}$CO do not appear to have strong corresponding emission in $^{13}$CO, we required that clumps have a $^{13}$CO signal-to-noise (S/N) ratio within the clump's $^{12}$CO boundaries of S/N $\geq$3 above 4$\sigma$ to apply the LTE method.  Our noise estimate was derived from the RMS noise in the Southern half of the cube ($\delta \lesssim -73^\circ$35') where no strong CO emission was detected in either line. 

Only 29 of the 110 clumps (26\%) fulfill this requirement, while the remaining 81 clumps (74\%) do not possess any strong $^{13}$CO emission.  To overcome this obstacle, our mass estimation method takes place in two parts (described in this and the following subsection \S\ref{subsec:xfactor}) and is similar to the approach taken by \citet{WongOudshoorn2022}, who in their study of the LMC's 30 Doradus region found that 53\% of CO-detected clumps were only traced by $^{12}$CO and did not have corresponding $^{13}$CO emission. 
The majority of the clumps in N90 with correponding $^{13}$CO emission are located in the NE rim and non-rim sections near NGC 602, with several others on the NW rim near the massive O3 star Sk 183 (see definitions of these subregions in Figure \ref{fig:spatial_props}a). 

We assumed that $^{12}$CO is optically thick and that its excitation temperature $T_{ex}$ is a function of brightness temperature,
\begin{equation}
T_{ex} = \frac{11.1 \rm{K}}{\ln(\frac{11.1}{I_{12}+0.19}+1)},
\end{equation}
where $I_{12}$ is the $^{12}$CO(2--1) intensity in K.   Since this result only relies on $^{12}$CO emission, we derived this quantity for all clumps.  Calculated $T_{ex}$ range from 6.5--28 K (Figure ~\ref{fig:spatial_props}c).  

We assumed that \thirtCO is optically thin, that \thirtCO and \twelveCO share the same $T_{ex}$, and that their relative abundance is constant.  We found the \thirtCO(2--1) optical depth of each PPV pixel for the 29 $^{13}$CO-traced clumps as \citep{GardenHayashi1991,bourke1997, indebetouw_alma_2013, wong_alma_2017},
\begin{equation}
    \tau_{0}^{13} = -\ln\left[1 - \frac{T_{B}^{13}}{10.6} \left\{\frac{1}{e^{10.6/T_{ex}}-1} - \frac{1}{e^{10.6/2.7}-1} \right\}^{-1} \right],     
\end{equation}
and column density N($^{13}$CO) as
\begin{equation}
    \rm{N}(^{13}CO) = 1.2 \times 10^{14} \frac{(T_{ex} + 0.88 \ K)\ e^{5.29/T_{ex}}}{1-e^{-10.6/T_{ex}}} \int\tau_{\nu}^{13}d\nu. 
\end{equation}
The maximum N($^{13}$CO) was $2.6 \times 10^{16}$ cm$^{-2}$.  We assumed an abundance ratio of H$_{2}$ to \thirtCO of $1.25 \times 10^{6}$ following \citet{jameson_first_2018} for the SMC, such that N(H$_2) = 1.25 \times 10^{6}$  N($^{13}$CO).  This abundance ratio is the combination of $^{12}$C/H and $^{12}$CO/$^{13}$CO abundance ratios; the former is constrained by UV absorption measurements \citep[see references in][]{TchernyshyovMeixner2015} to $\pm$40\%, and the latter by NLTE modeling of CO emission lines at $\sim$10pc resolution \citep{Nikolic2007}, with another 40\% uncertainty.  The total $^{13}$CO/H$_2$ could then be off by a factor of two, most likely in the direction that underestimates H$_2$.  The maximum H$_{2}$ column density observed was $3.2 \times 10^{22}$ cm$^{-2}$.

We calculated LTE masses for the clumps with significant $^{13}$CO detections as 
\begin{equation}
    M_{LTE} = 1.36 \ m_{H_2} \sum \rm{N(H}_2) \delta x \delta y,
\end{equation}
where 1.36 is a factor derived from cosmic abundances to convert from H$_2$ mass to total mass including helium, $\delta$x and $\delta$y are pixel sizes, and $m_{H_2}$ is the mass of an H$_2$ molecule.  LTE clump masses ranged from 13 $M_{\odot}$ to 286 $M_{\odot}$, with a median mass of 64 $M_{\odot}$ and total mass of all $^{13}$CO-traced clumps of $2435 \pm 330 \ M_{\odot}$.

Of course, it is possible that the clumps are not in LTE.  The systematic effects of using the LTE approximation can be understood by analyzing many non-LTE models.  From large grids of Radex \citep{vanderTakBlack2007} models, we find that under typical molecular cloud conditions, where $^{12}$CO has optical depths of a few and $^{13}$CO between $\sim$0.5–2, the LTE method tends to slightly (10-20\%) overestimate the $^{12}$CO excitation temperature (since its optical depth is less than the infinite assumed).  On the other hand, if $^{13}$CO has optical depth $<$1, its excitation temperature can be lower than that of $^{12}$CO by up to a factor of 2.   For very cold clouds ($T_K\lesssim$10 K), the LTE method underestimates the true $^{13}$CO column density by up to a factor of $\sim$2, and for very dense ($n_H\gtrsim$5000 cm$^{-3}$) and warm ($T_K\gtrsim$50 K) clouds, the LTE method overestimates the $^{13}$CO column density by up to a factor of $\sim$2.  However, the calculated $^{13}$CO column density from the LTE method is within 25\% of the true value for fairly wide ranges of parameter space: $^{12}$CO column densities between 10$^{16}$ and 3$\times$10$^{18}$, and 15 K$\lesssim$T$_K\lesssim$65 K.  For the N90 clumps we apply the LTE method to, the average N($^{12}$CO) is 5.9$\times$10$^{17}$ and the average $T_{ex}$ is 19 K, so we expect the LTE mass estimates to be reasonable.

\vspace{1cm}
%%%%%%%%%%%%%%%%%%%%%%%%%%%%%%%%%%%%%
\subsubsection{CO-to-H$_2$ Conversion Factor} \label{subsec:xfactor}

\begin{figure}
    \centering
    \includegraphics[width=0.45\textwidth]{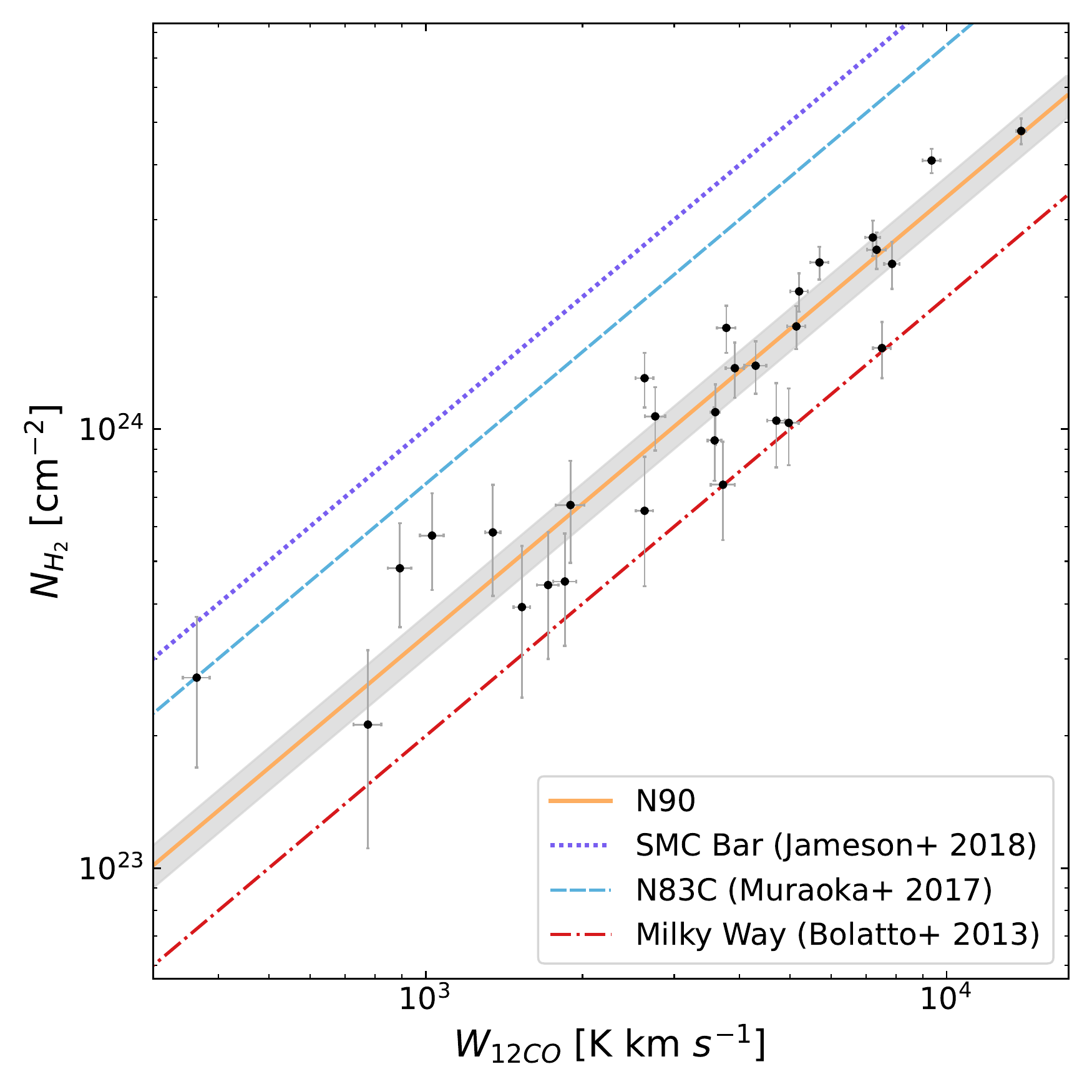}
    \caption{Total column density N(H$_2$) is compared to total integrated $^{12}$CO intensity $W_{12CO}$ for clumps with $^{13}$CO S/N $\geq$ 3.  The solid orange line shows the best fit CO-to-H$_2$ conversion factor $X_{CO,B}$ for the clumps in N90, $X_{CO,B}=(3.4 \pm 0.2) \times 10^{20}$ cm$^{-2}$ (K km s$^{-1}$)$^{-1}$, and is surrounded by gray shading showing the $95\%$ confidence interval for the fit.
    The dotted purple line was fit to star forming regions in the SMC Bar by \citet{jameson_first_2018}, with $X_{CO}=1 \times 10^{21}$ cm$^{-2}$ (K km s$^{-1}$)$^{-1}$ .  The dashed blue line was fit to clumps in N83C in the SMC Wing by \citet{muraoka_alma_2017}, with $X_{CO}=7.5 \times 10^{20}$ cm$^{-2}$ (K km s$^{-1}$)$^{-1}$.  The dot-dashed red line shows a typical  $X_{CO}=2 \times 10^{20}$ cm$^{-2}$ (K km s$^{-1}$)$^{-1}$ for the Milky Way \citep{bolatto_co--h2_2013}.}
    \label{fig:xfac_w12co}
\end{figure}

Although the LTE method described above is powerful when multiple lines are traced, it is limited and becomes less reliable for clumps with weak $^{13}$CO detections.  To circumvent this issue, we derived an $X_{\textrm{CO}}$ CO-to-H$_2$ conversion factor from the clumps traced by $^{13}$CO, and applied it to the 81 clumps with $^{13}$CO S/N $\geq$ 3 to obtain estimates of their masses.  We use the notation $X_{CO,B}$ to indicate that this factor is only intended to include gas that is $^{12}$CO-``bright'', and does not account for diffuse ``CO-dark'' gas in clump envelopes.  We discuss the role of CO-dark gas in N90 in \S\ref{subsec:dark}.

We fit $X_{CO, B}$ as
\begin{equation}
\rm{N}(\rm{H}_2) = X_{CO, B} \ W_{12\rm{CO}}
\end{equation}
where N(H$_2$) is the total H$_2$ column density in units of cm$^{-2}$ and W$_{12\rm{CO}}$ is the integrated $^{12}$CO line intensity in units of K km s$^{-1}$.  In the Milky Way, \citet{bolatto_co--h2_2013} recommended an average value of $X_{\textrm{CO, MW}} = \ 2 \times 10^{20}$ cm$^{-2}$ (K km s$^{-1}$)$^{-1}$.  We performed ordinary least squares regression with a heteroskedasticity-consistent standard error estimator (``HC3'' in the python package {\tt statsmodels}, \citet{statsmodels_seabold2010}).  The best-fit slope was $X_{\textrm{CO,B}}=(3.4\ \pm \ 0.2) \times 10^{20}$ cm$^{-2}$ (K km s$^{-1}$)$^{-1}$, or equivalently $X_{\textrm{CO,B}} \sim 1.7 \ X_{\textrm{CO, MW}}$.  

This fit is shown in Figure \ref{fig:xfac_w12co}. The 95\% confidence interval for $X_{\textrm{CO,B}}$ is [3.0 $\times$ 10$^{20}$, 3.75 $\times$ 10$^{20}$] cm$^{-2}$ (K km s$^{-1}$)$^{-1}$, and standard fit diagnostics suggest this model is adequate (coefficient of determination $R^2 = 0.95$, $F$-test statistic calculated for robust covariance of $F=309.8$ with $p < 0.001$).  
The $X_{\textrm{CO,B}} \sim 1.7 \ X_{CO,MW}$ we derived is consistent with, albeit slightly lower than, values of $X_{\textrm{CO,B}}$ found in other SMC regions observed at parsec-scales in $^{12}$CO and/or $^{13}$CO:  
In the Magellanic Bridge, \citet{kalari_resolved_2020} and \citet{valdivia-mena_alma_2020} derived values of $X_{\rm{CO}}$ of $\sim$2--4 $X_{\textrm{CO,MW}}$, while  \citet{muraoka_alma_2017} found $X_{\textrm{CO}}\sim 4 \ X_{\textrm{CO,MW}}$ in the star forming region N83C in the southeast Wing.  \citet{jameson_first_2018} similarly found an average $X_{\textrm{CO}}\sim 5 \ X_{\textrm{CO,MW}}$ across several star forming regions in the Southwest Bar of the SMC at $<$3 pc scales.

We applied our derived $X_{CO,B}$ factor to the clumps without significant $^{13}$CO detections and derived masses $M_{X_{CO,B}}$ ranging between 1.2 -- 52 M$_\odot$, with a mean value of 10.7 M$_\odot$ and total mass of 860 M$_\odot$.  We also applied this factor to the clumps traced by $^{13}$CO.  Going forward, we use the $M_{X_{CO,B}}$ mass estimates for all clumps for consistency.  We find a total gas mass for all clumps traced by $^{12}$CO in N90 of M$_{\rm{X_{CO,B}}}$ $\sim 3310 \pm 250 \ M_{\odot}$.  The 26\% of clumps that are traced by $^{13}$CO contribute 74\% of the total CO-bright clump mass.     

Our mass estimate is slightly lower than the total clump mass estimate of 3800 $M_\odot$ in N90 made by \citet{fukui_formation_2020} using only the 7m $^{12}$CO ALMA observations.  The difference between these results stems from variations in mass calculation and clump identification methods, not in spatial filtering, because $>$95\% of the $^{12}$CO flux in the 7m map is recovered in our 12+7m map).  \citet{fukui_formation_2020} identified 19 clumps with radii between 1.9--3 pc and used an $X_{CO, B}$= 7.5$\times 10^{20}$ cm$^{-2}$ (K km/s)$^{-1}$ \citep{muraoka_alma_2017}, as opposed to our sample of 110 clumps with radii between 0.2--0.8 pc and lower adopted $X_{CO,B}$.

\subsubsection{Other Clump Properties}\label{S:clump_props}

A full catalog of clump properties is presented in machine-readable format in Table \ref{tab:clumps}.  We determined the radius $R$ of each clump by fitting an ellipse to its half-light contour and converting the FWHM values of the ellipse’s major and minor axes to the standard deviation of a Gaussian profile.  We then multiplied by 1.91 to calculate the ``effective radius'' as defined by \citet{solomon_mass_1987}.  We report radii as the geometric mean of the major and minor axes, and these values range from 0.26 pc to 0.77 pc with a median of 0.40 pc.

We calculate average surface densities as $\Sigma = M_{X_{CO,B}} / (\pi R^2)$.  The median $\Sigma$ is 24 $M_\odot$ pc$^{-2}$, with a standard deviation of 40 $M_\odot$ pc$^{-2}$.  To estimate the volume densities of clumps, we assume that the clumps follow a power-law density profile,
\begin{equation}
    \rho(r) = \rho_c \ \left(\frac{r}{R_{0}}\right)^{-k},
    \label{eqn:density_prof}.
\end{equation}
where $R_0$ is a normalizing radius that we set to be $R_0 = 0.1$ pc for all clumps, and $\rho_c$ is the density of the clump at $R_0$.  Using the derivation in Appendix A.1 of \citet{ONeill2022_codark}, this central density can be estimated as
\begin{equation}
\rho_c = \frac{(3-k)}{4\pi} \frac{M(r)}{R_0^k \  r^{3-k}}
\label{eqn:rhoc}
\end{equation}
Power-law indices of $k\sim$1--2 have frequently been derived for clumps and cores \citep[e.g.,][]{caselli2002, pirogov09,chen2019,chen2020,lin2021} and we adopt $k=1$ for the N90 clumps.  Through solving Eqn. \ref{eqn:rhoc} with $r=R_{CO}$, densities at 0.1 pc range between 10$^1$--10$^3$ M$_\odot$ pc$^{-3}$, with an average of $\rho_c \simeq 190$ M$_\odot$ pc$^{-3}$. {\color{red} }

We calculated velocity dispersions $\sigma_{v}$ and peak CO velocities $v_{LSRK}$ by fitting Gaussian distributions to intensity-weighted $^{12}$CO velocity profiles.  We do not correct $\sigma_v$ for the expected contribution from thermal motion ($\sim$0.08 km s$^{-1}$ for CO at 20 K); this is discussed further in \S\ref{subsec:sizeline}.  Values of $\sigma_{v}$ ranged from 0.23 km s$^{-1}$ to 1.07 km s$^{-1}$ with a median of 0.45 km s$^{-1}$.  Peak $^{12}$CO velocities ranged between $v_{LSRK} =$159--179 km s$^{-1}$, with a mean of 167 km s$^{-1}$.

%%%%%%%%%%%%%%%%%%%%%%%%%%%%%%%%%%%%%%%%%%%%%%%%

%%%%%%%%%%%%%%%%%%%%%%
\subsection{Archival Data: IR-identified YSO candidates}\label{sec:archival}
\begin{figure}
    \centering
    \includegraphics[width=0.47\textwidth]{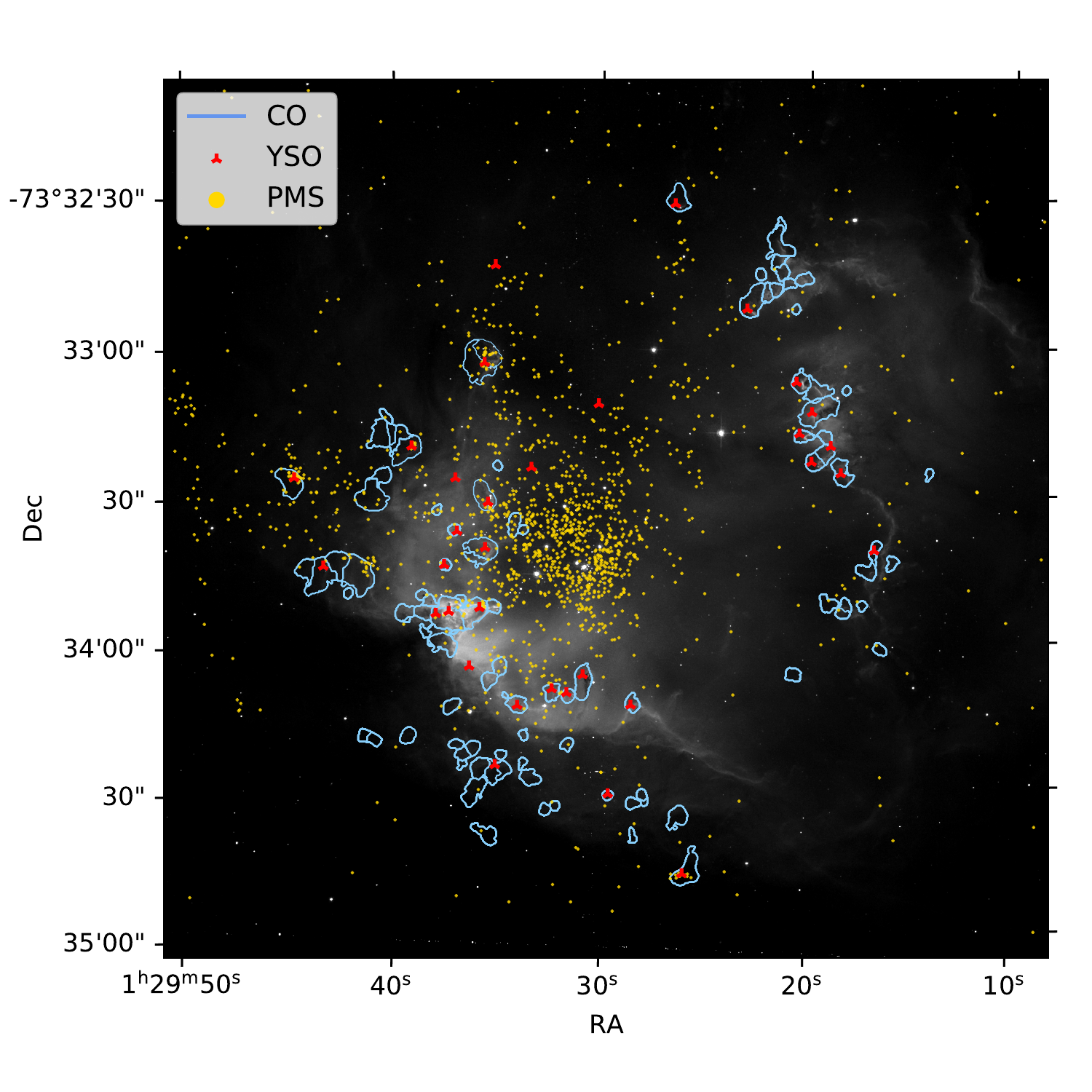}
    \caption{The HST ACS H$\alpha$ image is shown with the locations of PMS stars (yellow points), YSO candidates (red triangles), and the 2D boundaries of the CO clumps identified by the {\tt quickclump} algorithm (light blue contours).}
    \label{fig:pms_halpha}
\end{figure}

Two types of young stellar objects have been analyzed in the N90 region: solar-mass PMS stars identified with high resolution HST optical and near-IR photometry, and YSO candidates with infrared excess emission attributed to circumstellar dust identified using Spitzer and Herschel Space Telescope (Herschel) data. 
We revisit these populations here in order to assess their relationship with the resolved molecular gas, but do not attempt to re-do the careful classification of previous authors.

\subsubsection{HST Data}

\begin{figure}
    \centering
    \includegraphics[width=0.47\textwidth]{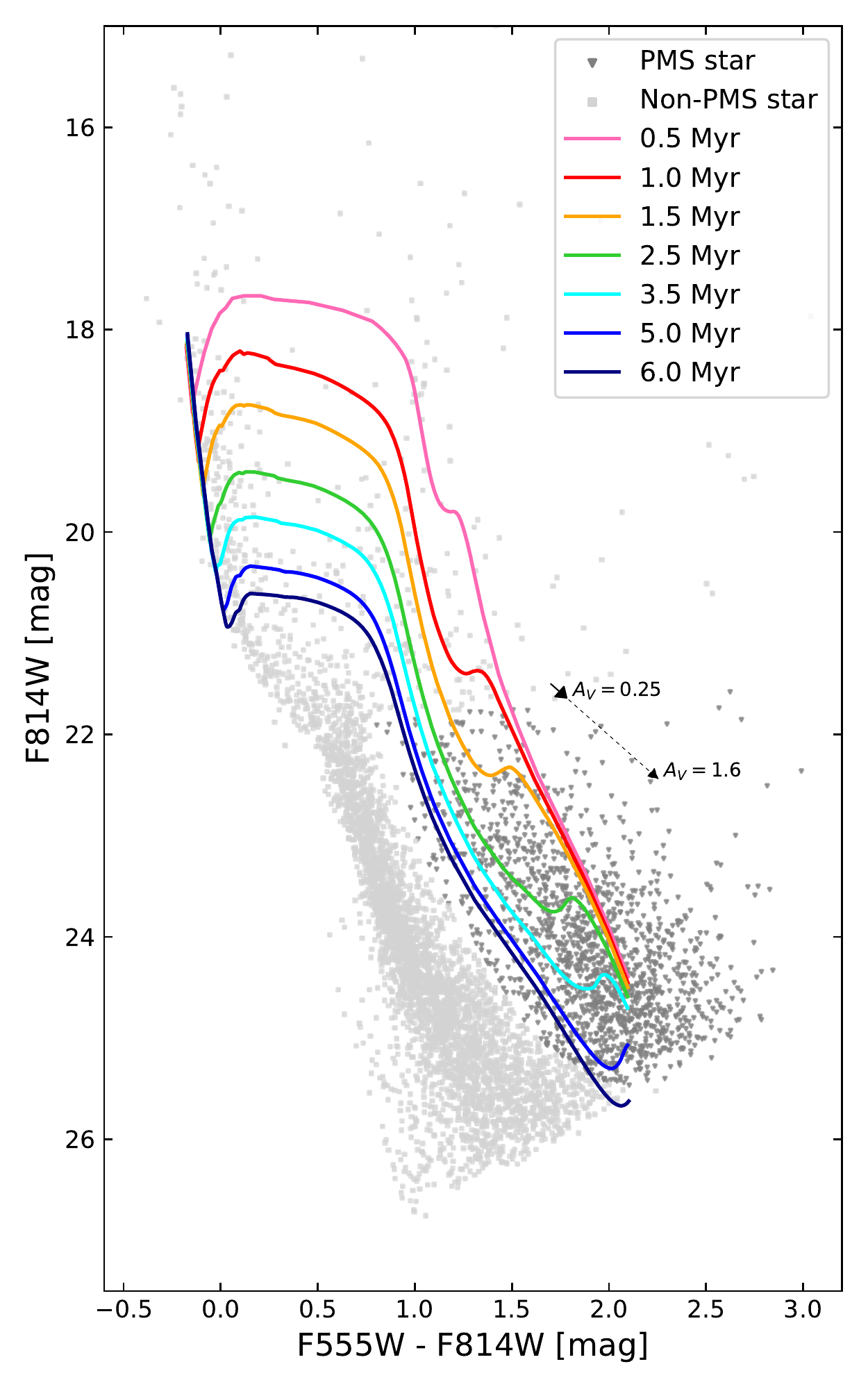}
    \caption{Locations of stars in F555W - F814W, F814W color-magnitude space.  PMS stars are marked with dark gray triangles, and non-PMS stars are marked with light grey squares.  PMS isochrone models for ages between 0.5 and 6 Myr are shown as colored curves. A reddening vector for $A_V=0.25$ mag is shown with by the thick black solid arrow, and for $A_V=1.6$ mag with the thin black dashed arrow.}
    \label{fig:pms_iso}
\end{figure}

We used HST F555W ($\sim$V) and F814W ($\sim$I) observations of N90 reduced by \citet{schmalzl_initial_2008} to study the distribution of solar-mass PMS stars in the region, as selected by \citet{gouliermis_clustered_2012} (herafter G12).  The locations of PMS stars are shown in Figure \ref{fig:pms_halpha}.  The data from HST GO program 10248 consist of 2156s and 2269s total integration time with ACS WFC in F555W and F814W, respectively. Photometry was performed using DOLPHOT\footnote{The ACS module of DOLPHOT is an adaptation of the photometry package HSTphot \citep{Dolphin2000}. It can be downloaded from \url{http://americano.dolphinsim.com/dolphot/}.}, reaching a depth of 26 mags in each filter, albeit at reduced completeness below 23 and 22.5 magnitudes in F555W and F814W, respectively \citep[see][for details]{schmalzl_initial_2008}. 

G12 separated faint PMS stars from lower main sequence stars by fitting pairs of Gaussians to the distribution of stars contained within bands perpendicular to the main sequence.  The minimum between the Gaussian corresponding to the MS population and that corresponding to the redder PMS population is taken as the classification boundary.  We select PMS stars as falling within or to the right of the region bounded by F814W > 21.8 mag and F814W < -1.9(F555W - F814W)$^2$ + 8.3(F555W - F814W) + 16.6.  We generated isochrones using the Pisa PMS evolutionary models \citep{Tognelli2011} and the IDL program TA-DA \citep{da_rio_ta-da_2012} to create synthetic photometry based on the \citet{Kurucz1993} atmospheric models.  The models were calculated for Z=0.003, Y=0.254, and mixing length parameter $\alpha$ = 1.2.  We assumed $A_V=0.25$ mag and E(B-V) = 0.08 mag \citep{carlson_progressive_2007,gouliermis_clustered_2012}, $R_V=3.1$ \citep{Schultz1975, GordonClayton2003}, and a distance modulus of $\mu = 18.91$ mag (corresponding to $\sim$60.6 kpc, \citealt{hilditch2005}).  We assigned ages to each PMS star based on the track it was closest to for ages between 0.5 and 6.5 Myr in 0.25 Myr steps. 

The resulting isochrones for ages of 0.5, 1, 1.5, 2.5, 3.5, 5, and 6 Myr are shown in Figure \ref{fig:pms_iso} with the selected PMS stars.  In \S\ref{S:pms}, we estimate an average $A_V \simeq 3.1$ mag through the centers of CO clumps that contain PMS stars.  For the $\sim$10\% of PMS stars that appear contained within projected 2D clump, a typical embedded star in the center of a clump might then reasonably be affected by half of this value, $A_V \simeq 1.6$ mag.  Reddening vectors for both $A_V$=0.25 mag and $A_V$=1.6 mag are plotted in Figure \ref{fig:pms_iso}.  They fall at steep angles to the isochrones and indicate that high levels of differential reddening could significantly skew age estimates.  We discuss this possibility further in \S\ref{S:pms}. 

%%%%%%%%%%%%%%%%%%%%%%%%%%%%%%%%%%%%%%%%%%%%%%%%%%%%%%%%%

\subsubsection{Spitzer and Herschel Data}

We also re-analyzed Spitzer-identified intermediate- and high-mass YSO candidates in the region.  \citet[][herafter C11]{carlson_panchromatic_2011} combined V, I, J, H, K, Spitzer IRAC 3.6-8.0 $\mu$m, and Spitzer MIPS 24 $\mu$m photometry, using the high-resolution optical data to remove background galaxies.  C11 first fit the sources with stellar photospheres, removing sources consistent with stars, and then fit the remainder with \citet{Robitaille2006} YSO models to identify and classify all sources consistent with intermediate-mass YSOs.
Starting with the C11 combined photometry catalog with galaxies removed, but including sources they classified as stars, we added aperture photometry from the Herschel HERITAGE survey \citep{Meixner2013,Seale2014} at 100, 160, 250, and 350 $\mu$m using the Spectral and Photometric Imaging Receiver (SPIRE) and Photodetector Array Camera and Spectrometer (PACS). 

We re-calculated aperture photometry of all Spitzer and 2MASS images in order to directly compare our aperture photometry code against the catalog photometry, and have a consistently calculated number for all bands in which an upper limit was required. Our script simply extracts the pixel sum in circular apertures at the source location, with radii of [1, 1,  3, 3, 3, 3,  9, 18, 8, 11, 18, 25, 37] arcseconds in filters [F555W, F814W, IRAC1, IRAC2, IRAC3, IRAC4, MIPS24, MIPS70, PACS100, PACS160, SPIRE250, SPIRE350, SPIRE500].  A background consisting of the median value in an annulus around each aperture was subtracted.  
The aperture photometry agrees within uncertainties with the previous C11 photometry for most sources, except those that suffer from confusion and crowding. 

We visually assessed the spectral energy distribution (SED) and image cutouts in all filters using a script to assemble that information on a single page for each source, in order to evaluate which filters were contaminated by neighboring sources and/or diffuse emission; in the case of contamination, we used the aperture flux density as an upper limit in subsequent fitting.  After this assessment, most of the sources had to be fit with upper limits in the new longer-wavelength bands because of the lower angular resolution of those data; however, these upper limits are still sufficient to exclude some models included in C11 that are very bright in the far-infrared.

We fit the sources with the updated \citet{Robitaille2017} set of YSO models, which cover a wider range of parameter space more uniformly than the previous \citet{Robitaille2006} grid. However, the newer models do not have associated stellar masses as the 2006 models did, so we match each model's log L and log T to the nearest PARSEC PMS photosphere model \citep{Bressan2012} to determine an $M_\star$.  The new set of models are parameterized by envelope characteristic density $\rho_0$ and centrifugal radius $R_C$.  Given that the circumstellar envelope has the particular density distribution of a rotating infalling toroid, one can uniquely calculate an "envelope accretion rate" $\dot{M}$ for a model's given $\rho_0$, $R_C$, and stellar mass.  This parameter is largely a convenient way to parameterize the degree to which the source is embedded, with higher $\dot{M}$/$M_{\star}$ indicating a more embedded source as discussed in \citet{Robitaille2006}.  When we compare these properties to CO clumps in \S\ref{S:yso}, the general properties of the YSO candidate population will be considered, but the precise fitted mass or envelope mass will not dramatically alter our conclusions.

For each source, we calculate $\chi^2$ for each model, and use the probability-weighted mean value of each fit parameter and the full-width at half-maximum of the parameter's 1D marginalized probability density function (PDF) as the fitted parameter and its uncertainty.  We examined all sources' PDFs as a function of M$_\star$ and envelope $\dot{M}$ - a minority of sources have mulitply-peaked PDFs but the adopted uncertainty range in all cases encompasses both peaks so is a reasonable measure of the data's ability to constrain the source properties.

Our re-analysis does not change the list of intermediate-mass YSO candidates relative to C11.  The primary addition to C11's analysis is that the addition of longer-wavelength upper limits eliminates luminous, heavily embedded models, with high circumstellar dust columns.  Addition of longer-wavelength photometry in a few cases suggests an infrared excess for some sources classified by C11 as bare stellar photospheres, but none of these additions are definitive, and higher resolution long-wavelength imaging will be required to conclusively measure any infrared excess.

We preserve C11's classification of ``K'' source as non-YSOs in our analysis: K049 (J012903.28-733413.2) has a tentative 100 $\mu$m detection of 35$\pm$15 mJy, but is located in filamentary diffuse emission. K194 (J012920.73-733327.1) has a marginal 100 $\mu$m measurement of 4.5$\pm$3 mJy.  The most likely IR-excess candidate amongst C11's "K" sources is K456 (J012954.82-733231.5) with a marginal 100 $\mu$m measurement of 2.7$\pm$1.6 mJy, but also a 24 $\mu$m flux of 1.1$\pm$0.45 mJy in excess of a stellar photosphere.  All three lie outside of the region mapped in CO, so their classification has no effect on our conclusions. 
C11 noted two sources that they called stars but with infrared excess emission in their analysis with longest wavelength of 24 $\mu$m.  S235 and S213 are relatively brights star located within the central bubble; we confirm that both have 24 $\mu$m emission in excess of a photosphere, but neither are conclusively detected at longer wavelengths, as they would be if they had a massive circumstellar envelope, so we keep the C11 "star" classification.  Neither is associated with CO emission. 

The longer-wavelength data do, however, significantly change the stellar and envelope masses for the most massive sources - all of the most massive sources fit by C11 are fit with models that have bright FIR excess emission, and in all cases, we measured the FIR emission to be modest, excluding those massive models that are acceptable fits to C11's shorter wavelength range.  The largest fit YSO mass is 8$\pm$1 M$_\odot$, whereas C11's best fits included one object consistent with 26 M$_\odot$ (we find 4$\pm$1 M$_\odot$) and five with masses between 10 and 12 M$_\odot$.  In total, we find a higher mass estimate than C11 for only one YSO (Y227, J012937.37-733352.4, with C11 mass of 6.86 M$_\odot$ vs our estimate of 6.99 M$_\odot$); all other YSO candidates have lower masses.  We derived a total YSO mass of $\sim$160$ M_{\odot}$ in N90, which is significantly less than C11's previous estimate of a total YSO mass  of $\sim$300 $M_{\odot}$.  The fit properties of all YSOs are reported in Table \ref{tab:yso}.  

Our fits to the Robitaille models yield two constraints on the age of the intermediate-mass YSO candidates:  the age of the PARSEC photospheric model matched with each \citet{Robitaille2017} YSO model, and the fitted mass divided by fitted envelope accretion rate.  The “photospheric age” is fairly well constrained, to $\sim$10$^5$ years for the more massive YSOs $>$5 $M_\odot$, increasing to $\sim$10$^6$yrs for the lower mass M$_\star\simeq$ 2 M$_\odot$ YSOs. 
The fitted mass divided by accretion rate is consistent, agreeing to within the order-of-magnitude uncertainties on that ratio, but has the even larger caveat that the current fitted accretion rate is almost certainly not the accretion rate through the entire mass assembly of the YSO, and that the fitted accretion rate only has physical meaning insofar as the rotating infalling envelope density distribution used by Robitaille actually represents the accretion process - in general we are far more confident that this fitting process can constrain the total envelope dust mass than we are in interpreting that quantity as an accretion rate.   Nevertheless, these two constraints suggest ongoing intermediate-mass star formation in NGC 602 over the last 1--2 Myr.

%%%%%%%%%%%%%%%%%%%%%%%%%%%%%%%%%%%%%%%%%%%%%%%%%%%%%%%%%%%%%%%%%%%%%%%%%%%%%%%%%%%%%%%%%%%%%%%%%
\section{Dynamics and Evolution of N90} \label{sec:structure}

%%%%%%%%%%%%%%%%%%%%%%%%%%%%%%%%%%%%%%%%%%%%%%%%%%%%%%%%%
\subsection{Morphology of N90}

\begin{figure}
    \centering
    \includegraphics[width=0.45\textwidth]{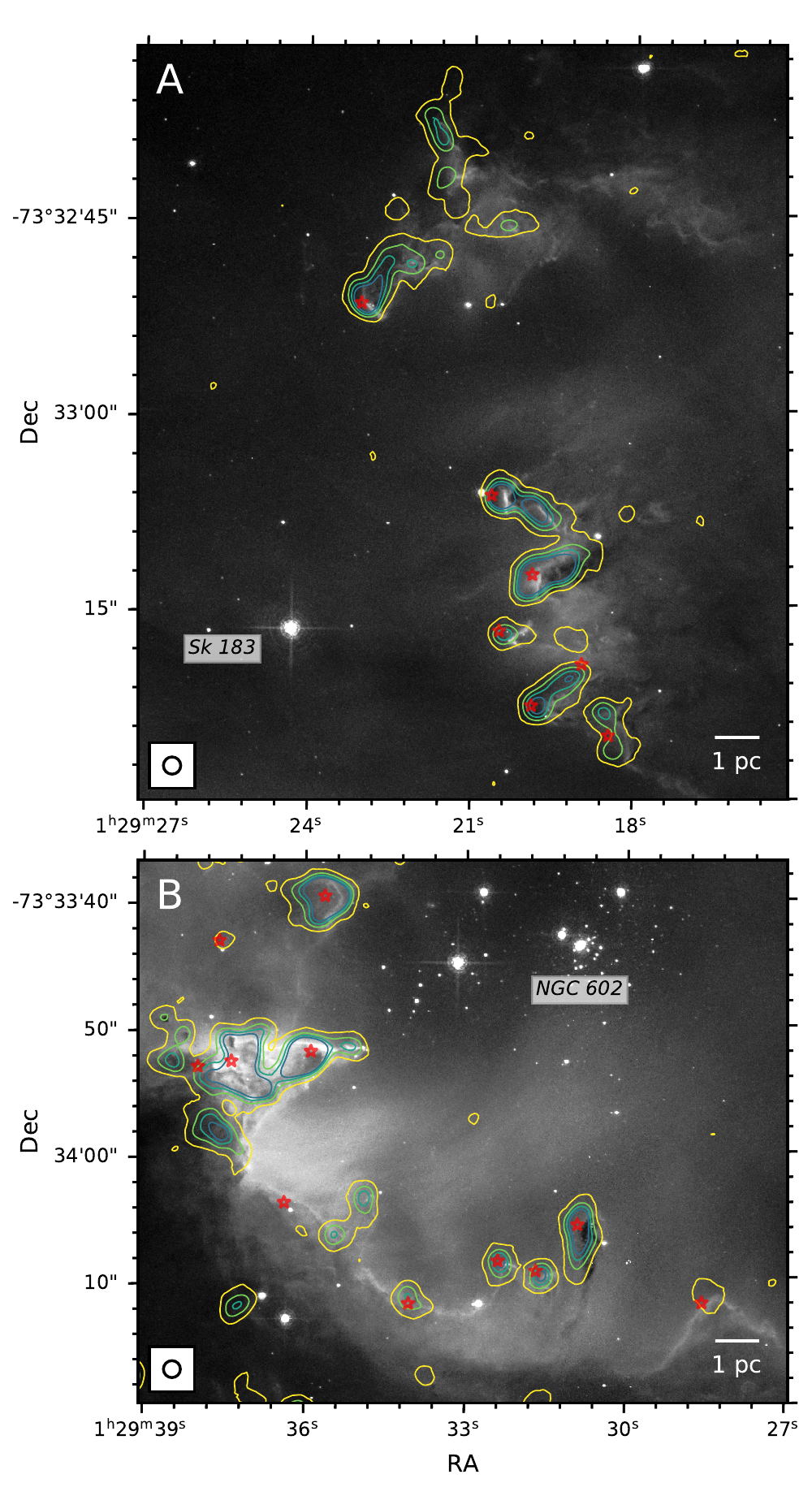}
    \caption{\textit{Top:} Expanded image of Figure \ref{fig:spatial_props}a's Region A, along the West rim near the massive star Sk 183 (labeled).  Contours of integrated \twelveCO intensity are shown, with levels of 0.25, 0.5, 0.75, and 1 Jy beam$^{-1}$ km s$^{-1}$ drawn in yellow, green, teal, and blue, respectively.  The synthesized beam size for the CO observations is shown in the lower left corner.  The locations of YSO candidates are shown by red stars.  HST ACS H$\alpha$ image is shown in gray.  \textit{Bottom:} As the left panel, but for Figure \ref{fig:spatial_props}a's Region B, on the East rim near NGC 602 (labeled).}
    \label{fig:pillars}
\end{figure}

N90 is characterized by a ring-shaped ``rim'' $\sim$30 pc in diameter that frames a central cavity containing NGC 602 (Figure \ref{fig:HubbleNGC602}b).  The rim is most clearly visible to the east and west, with the northeast section of the rim appearing to be more diffuse and the southeast section denser.   
Many background galaxies are clearly visible outside the edges of N90 due to the low surface density, generally transparent surrounding environment of the SMC Wing ($A_V\lesssim$10$^{-2}$ on degree scales; \citealt{gordon_dust_2014}).

N90's rim displays clear evidence of photodissociation and has many ``pillar''-like features.  Most of the molecular clumps in N90 are arranged along or immediately outside of the rim, and many of the pillars are closely associated with CO emission.  Two representative regions of the rim are shown in Figure \ref{fig:pillars}.  CO emission closely traces the edges of the H$\alpha$ emission, and most of the pillars point towards the center of N90; this suggests that they are the result of radiation from the cluster NGC 602 \citep{gritschneder_detailed_2010}.

Many of the YSO candidates in N90 (\S\ref{S:yso}) are embedded within the pillars and clumps along the rim.  In their study of $\sim$200 giant molecular clouds (GMCs) in the LMC, \citet{ochsendorf_location_2016} found that massive star formation is most likely to occur at the edges of clouds nearest to young stellar clusters, implying these clusters stimulate clump and star formation.  The YSO candidates in N90 shown in Figure \ref{fig:pillars} are also mostly concentrated on the edges of the rim/their host pillars that are closest to NGC 602 or Sk 183; this is especially apparent in Region A.

We assess whether the formation of the YSOs in N90 may have been triggered by NGC 602.  Here we define triggered star formation as a process in which the interaction between the earliest generation of stars and their environment directly causes the formation of subsequent generations.  This stands in contrast to sequential star formation scenarios where the formation of distinct stellar generations are largely unrelated.

\subsection{Spatial Variation in Clump Properties}\label{S:spatial}

\begin{figure*}
    \centering
    \includegraphics[width=\textwidth]{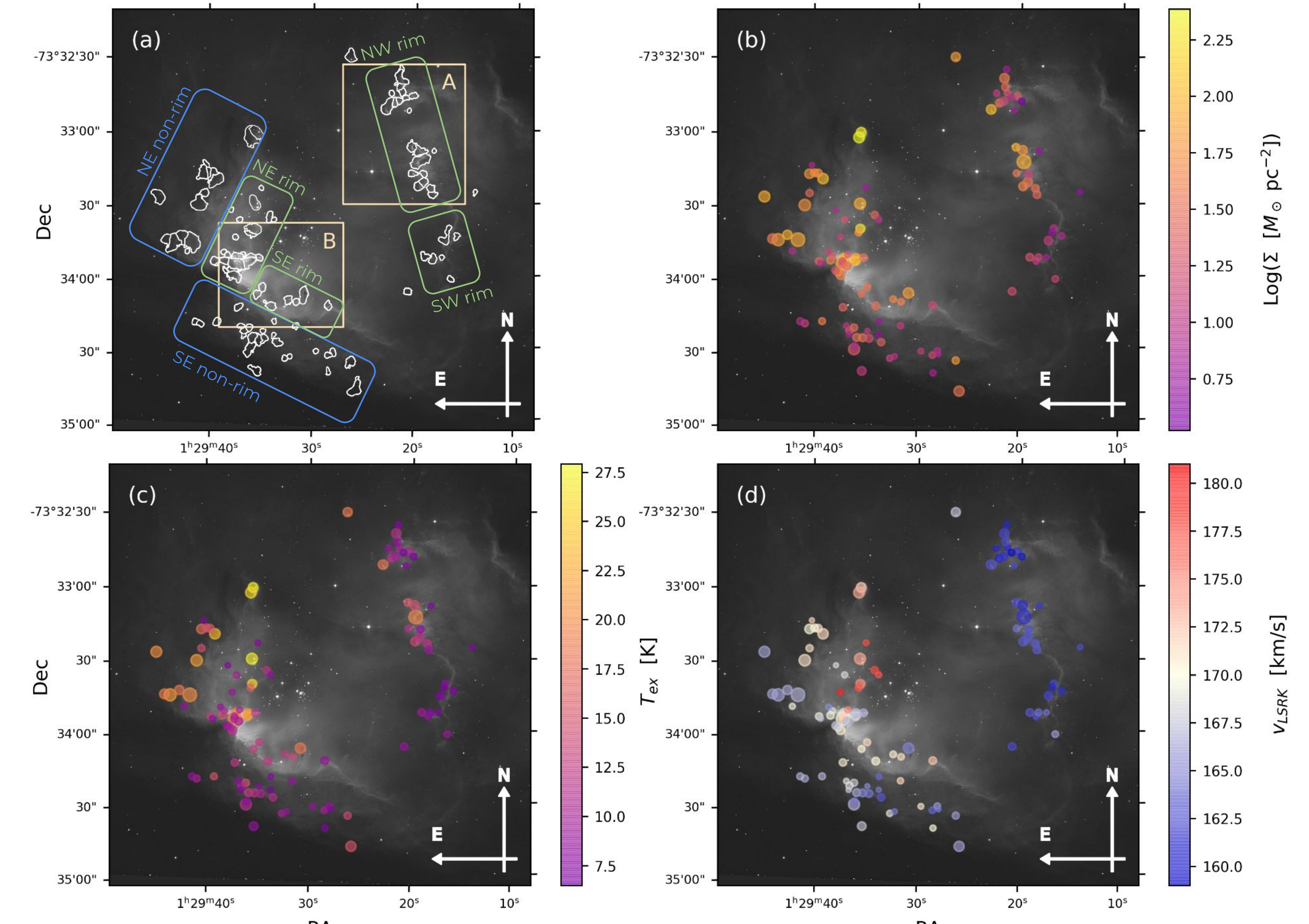}
    \caption{Spatial variation of CO clump characteristics, with HST H$\alpha$ image shown in gray.  \textit{(a:)} The regions we define as part of the rim are shown in green, and the corresponding non-rim regions are marked in blue.  The 2D clump boundaries identified by the {\tt quickclump} algorithm are shown in white.  Details of the two orange rectangles A and B are shown in Figure \ref{fig:pillars}.  \textit{(b:)} Average surface density $\Sigma$ of clumps, with point sizes proportional to clump area.  \textit{(c:)} Maximum clump excitation temperature $T_{ex}$.  \textit{(d:)} Peak $^{12}$CO velocity of clumps, with red/blue color scale centered at the mean radial velocity of \HI{} components surrounding N90, $v_{LSRK} \sim$170 km s$^{-1}$, as derived from \citet{fukui_formation_2020} and \citet{nigra_ngc_2008}.}
    \label{fig:spatial_props}
\end{figure*}

\begin{figure}
    \centering
    \includegraphics[width=0.44\textwidth]{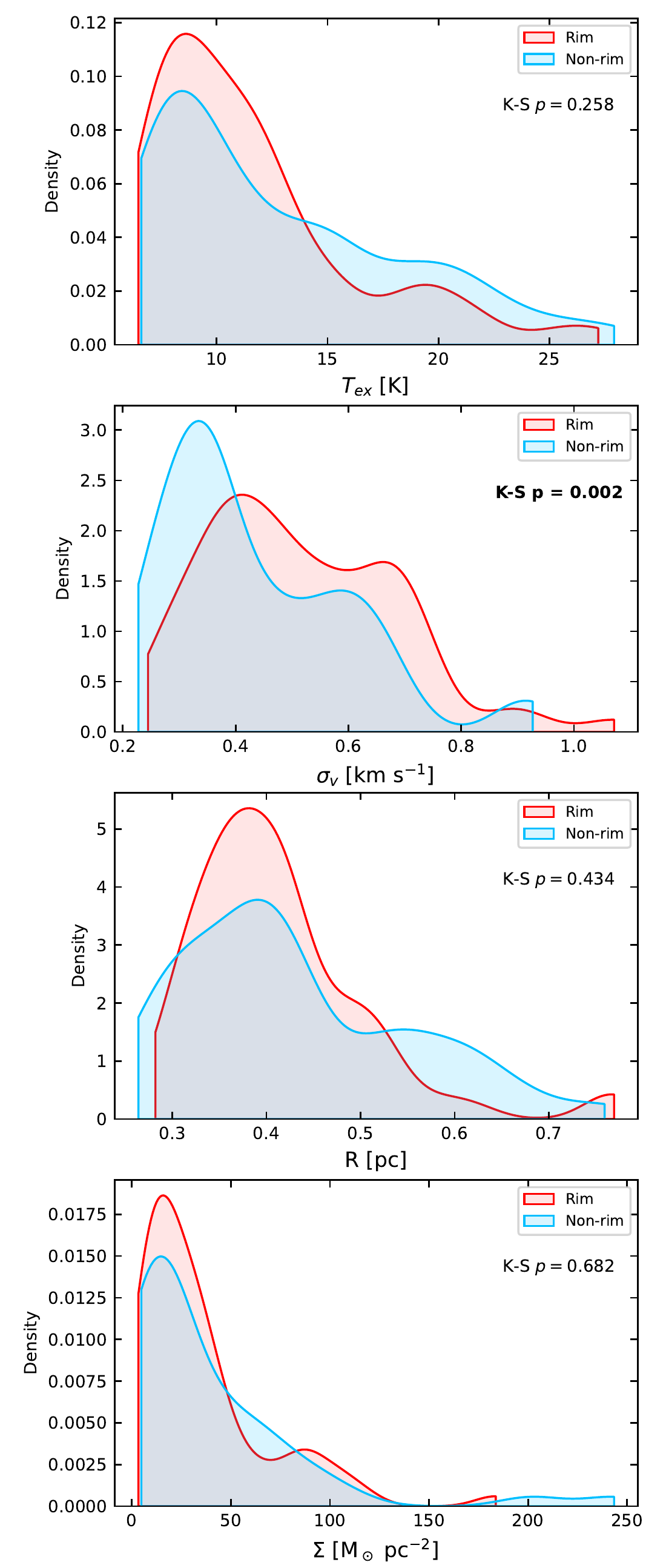}
    \caption{Kernel density estimations (KDEs) comparing physical properties of clumps on the N90 rim (red) and not on the N90 rim (blue), with both groups assigned as in Figure \ref{fig:spatial_props}a.  $p$-values resulting from two-sided K-S tests comparing the rim vs non-rim distributions are shown in each panel, with any $p < 0.05$ set in bold to indicate that there is a statistically significant difference between the distributions.  Top: Excitation temperature $T_{ex}$.  Second from top: Velocity dispersion $\sigma_v$.  Third from top: Radius $R$.  Bottom: Surface density $\Sigma$.}
    \label{fig:kde_rimprop}
\end{figure}

We assessed visually which CO clumps are on the rim (Figure \ref{fig:spatial_props}a) to investigate if any trends between clump properties and 2D position relative to the rim exist.  Through this assessment, we estimate that 64 of the clumps are on the rim, and 46 are outside of it.  76 of the clumps are located in the Eastern half of the region nearer to the NGC 602 cluster, and 34 are in the Western half nearer to the massive star Sk 183.  The spatial variation of the column densities, excitation temperatures, and radial velocities of the CO clumps are shown in Figure \ref{fig:spatial_props}b--d.  

Clumps in the NE, both in the rim and non-rim sections, appear to have higher surface densities (Figure \ref{fig:spatial_props}b) and excitation temperatures (Figure \ref{fig:spatial_props}c, which under the LTE assumptions is indirectly equivalent to \twelveCO brightness temperature) than clumps along the SE rim and non-rim sections.  The NE clumps may thus be more directly affected by the central OB association NGC 602, as gas associated with the H$\alpha$ rim could present less of a barrier to mechanical and radiative heating from the association than for clumps shielded by (and possibly located within or behind) dense gas to the south.  We used 2D radial distances between the center of NGC 602 ($\alpha=$01 29 32.133, $\delta=$ -73 33 38.13) and all clumps on the Eastern half of N90 as a proxy for the true 3D distances between their positions. We calculated Spearman's rank correlation coefficient $\rho_s$ between the distance from NGC 602 and clump properties; there are weak-to-moderate negative correlations between distance and $\sigma_v$ ($\rho_s = -0.40$, $p<0.001$) and distance and $v_{LSRK}$ ($\rho_s = -0.58$, $p<0.001$), and only weak correlations between distance and $R$, $M_{X_{CO,B}}$, or $T_{ex}$ (all $|\rho_s| < 0.3$).

Along the western rim, the warmest and densest clumps are concentrated in the NW section near the massive O3 star Sk 183 ($\alpha=$01 29 24.6, $\delta=$ -73 33 16.43).  Although the clumps in the northwest are relatively removed from NGC 602 itself, their proximity to Sk 183 could be responsible for their marginally higher excitation temperatures relative to the cooler and more isolated clumps in the SW rim. \citet{evans_rare_2012} found that Sk 183 is likely responsible for the majority of hydrogen-ionizing photons in N90 and \citet{ramachandran_testing_2019} confirmed its outsized effect, with Sk 183 is contributing up to 30\% of all ionizing photon flux in their sample of $\sim$300 OB stars scattered throughout the entirety of SGS 1.  We stacked the CO spectra within 5 pc of Sk 183 and identified no significant ($>$3$\sigma$) emission in either $^{12}$CO or $^{13}$CO.  This suggests that it has successfully dissociated its immediate surroundings, and could even be responsible for the nearby northern ``gap'' in the H$\alpha$ rim.  We also explored correlations between Western clump properties and their distance from Sk 183, and found moderate negative correlations between distance and: $\sigma_v$ ($\rho_s = -0.51$, $p=0.002$); $M_{X_{CO,B}}$ ($\rho_s = -0.48$, $p=0.004$); and $T_{ex}$ ($\rho_s = -0.49$, $p=0.003$).

Figure \ref{fig:kde_rimprop} compares the distributions of $T_{ex}$, $\sigma_v$, $R$, and $\Sigma$ between all rim and non-rim clumps.  The probability density functions of these properties are represented using Gaussian kernel density estimations (KDEs), with bandwidths chosen using ``Scott's Rule'' as implemented in the Scipy python package \citep{2020SciPy-NMeth}.  
By performing Kolmogorov-Smirnov (K-S) tests comparing the properties of rim vs non-rim clumps, we found no significant differences between the distributions of $R$ (with a $p$-value of $p=0.43$ at a significance level requirement of $\alpha = 0.05$), $T_{ex}$ ($p=0.26$), or $\Sigma$ ($p=0.682$) for the two groups.  There was a significant difference in the distributions of $\sigma_v$ between rim and non-rim clumps ($p=0.002$), with a one-sided Mann-Whitney U test revealing that clumps along the rim on average have larger $\sigma_v$ ($p=0.003$).

The overall lack of any strong correlations and relatively few significant differences between groups suggests that stellar feedback (e.g., radiation pressure or stellar winds) has not had a significant differential impact on the physical properties of the molecular clumps, or at least that any systematic trend cannot be extracted from the random cloud property variations with the modest range of central cluster distance available in these data ($\sim$2--23 pc).  \citet{indebetouw_alma_2013} found a similar lack of significant trends from 10 to 25 pc from the cluster R136 in the LMC region 30 Doradus.

\subsubsection{Clues for Formation of NGC 602}

We stacked the CO spectra in the half of the ALMA coverage to the south of N90 ($\delta \lesssim -73^\circ$35') where no clumps were identified by the {\tt quickclump} algorithm.  We found no significant ($> 3 \sigma$) CO emission in this area. Assuming $T_{ex} = 10$ K, the mean N($^{12}$CO) in this region is less than $4.5 \times 10^{16}$ cm$^{-2}$.  Thus, we find no strong evidence for CO to the south of N90 and conclude that CO-traceable molecular gas is largely localized in the site of massive star formation to the north, which is adjacent to the southern rim of SGS 1.  This correspondence suggests that turbulence and compression resulting from the southward expansion of \HI{} component(s) within SGS 1 triggered both the formation of dense molecular gas and then stars in N90.

\citet{fukui_formation_2020} proposed that the collision of two clouds hundreds of parsecs in diameter was responsible for the formation of NGC 602.  In this scenario, the larger of their two \HI{} clouds would have moved south from the northeast before colliding with a smaller, less-massive cloud moving north from the southwest.  They suggested that SGS 1 is simply a cavity inside the more massive cloud created by this collision. 
In the alternative colliding shells formation scenario proposed by \citet{nigra_ngc_2008}, one shell moved south from the northwest and the other moved south from the northeast, with NGC 602 forming at their intersection.

We examine if evidence of these collisions in SGS 1 7--8 Myr ago could still be preserved within the current clumps.  At solar metallicity, photoelectric heating rates in molecular gas are usually sufficiently higher than energy injection rates from the decay of turbulence that a relationship between dissipation of turbulence and kinetic temperature is not expected.  Mechanical heating can be traced by kinetic temperature when the heating is particularly high, as in central starbursts \citep[e.g.,][]{Kazandjian2016,Mangum2019}.  At low metallicity, however, both cooling rates and photoelectic heating rates are sufficiently reduced that even more modest levels of turbulent dissipation might be able to affect the gas kinetic temperature.  

Turbulence is expected to dissipate on the order of a crossing time \citep{Elmegreen2004}.  The HI clouds considered in \citet{fukui_formation_2020} have diameters d$\sim$ 600 pc, velocity dispersions $\sigma_v \sim$10 km s$^{-1}$, masses $M\sim 8 \times$ 10$^6$ M$_\odot$, and so a mean volume density $<$n$_H>$ of 3 H cm$^{-3}$.  The shell most closely associated with N90 by \citet{nigra_ngc_2008} has d$\sim$200 pc, $\sigma_v \sim$6 km s$^{-1}$, $M\sim$3$\times10^5$ M$_\odot$, and also $<$n$_H>$ = 3 H cm$^{-3}$.  Estimating a turbulent dissipation rate as 0.5<$\rho$>$\sigma_v^2$ / $\tau_{\rm{crossing}}$ with $\tau_{\rm{crossing}} \sim $d/$\sigma_v$ yields 1$\times 10^{-27}$ and 3$\times 10^{-27}$ erg s$^{-1}$ cm$^{-3}$ for the \citet{nigra_ngc_2008} and \citet{fukui_cloud-cloud_2021} clouds, respectively. The cooling rate for the neutral ISM dominated by C$^+$ and O$^0$ line emission at 1/5 $Z_\odot$ is $\lesssim$10$^{-26}$ erg s$^{-1}$ cm$^{-3}$ in clouds at these low densities \citep{wolfire_thermal_1995}. These estimates are sufficiently uncertain as to preclude a definitive statement, but the cooling rate exceeding the estimated heating rate suggests that any excess kinetic energy from the cloud collision has probably been radiated away, and feedback from the central cluster and Sk 183 is still more likely the dominant energetic driver. However it is not impossible that at low metallicity, the signature of more turbulent HI gas might still be present in current properties of the molecular gas formed from that HI gas.

We compare the radial velocities (RVs) of the clumps to nearby stars in N90 and to SGS 1's proposed \HI{} components.  All velocities are reported in the LSRK frame.  Clump RVs range from 161 to 179 km s$^{-1}$, with a mean of 168 km s$^{-1}$.   \citet{evans_rare_2012} derived an RV of $151 \pm 1$ km s$^{-1}$ for Sk 183 and \citet{ramachandran_testing_2019} derived a mean RV for $\sim$17 OB stars in N90 of $158 \pm 4$ km s$^{-1}$.  The cause of this offset from the RVs of the CO clumps is unknown. 
The \citet{ramachandran_testing_2019} measurements have typical uncertainties of $\pm 10$ km s$^{-1}$, making it difficult to tell how significant this shift is.  In contrast, the clump RVs are in close agreement with the RVs of nearby \HI{} structures along the ``ring'' of SGS 1: \citet{nigra_ngc_2008} derived an RV of $168 \pm 5$ km s$^{-1}$ for a proposed progenitor \HI{} gas clump ``curled'' around the south of N90.  Similarly, the larger of the two \citet{fukui_formation_2020} clouds had a range of velocities from 163 -- 183 km s$^{-1}$, with a peak at $\sim$173 km s$^{-1}$.  

In Fig. \ref{fig:spatial_props}d we show the spatial variation of the clump RVs relative to the mean of these two measurements, 170 km s$^{-1}$.  Clumps along the eastern rim near NGC 602 appear consistent with this value (median +0.2 km s$^{-1}$ from 170 km s$^{-1}$), while clumps in the northeast and southeast non-rim regions (median -3.1 km s$^{-1}$) and the western half (median -7.7 km s$^{-1}$) mostly appear blue-shifted.  \citet{Gvaramadze2021} derived a central RV of $177 \pm 6$ km s$^{-1}$ for background H$\alpha$ emission near SNR SXP 1062, which falls just outside of the SGS 1 \HI{} ``ring''.  Since these measurements of \HI{} and H$\alpha$ RVs along in SGS 1 are largely consistent with the radial velocities of the clumps in N90 on the edge of the ring, a close connection between SGS 1 and the formation of N90 appears likely.

%%%%%%%%%%%%%%%%%%%%%%%%%%%%%%%%%%%%%%%%%%%%%%%%%%%%%%%%%%%%%%%%%%%%%%%%%%%%%%%%%%%%%
\subsection{Clump Association with YSOs and PMS Stars} \label{subsec:formation}

Many previous studies have analyzed the characteristics of candidate PMS stars and YSOs in N90 \citep[e.g.,][]{carlson_progressive_2007,carlson_panchromatic_2011,cignoni_star_2009,schmalzl_initial_2008,gouliermis_clustered_2007,gouliermis_clustered_2012,de_marchi_photometric_2013}. To better inform our analysis of the observed CO clumps we replicate and extend some aspects of this extensive past analysis.
Hereafter, we refer to the candidate several solar-mass sources with MIR excess emission selected from Spitzer and Herschel as "YSOs", and the generally lower-mass sources selected from their location in an HST color-magnitude diagram as "PMS stars".

%%%%%%%%%%%%%%%%%%%%%%%%%%%%%%%%%%%%%%%%%%%%%%%%
\subsubsection{YSOs}\label{S:yso}

Of the thirty-three YSOs identified in \S\ref{sec:archival} that are inside the ALMA-observed area, twenty-eight are located inside the projected 2D boundaries of a CO clump ($\sim$85\%).  Correspondingly, 27 of the 110 clumps ($\sim$25\%) appear to contain at least one YSO.   By performing one-sided K-S tests, we found significant differences between the properties of clumps that contain vs. do not contain YSOs: clumps containing YSOs have higher $R$ ($p<0.001$), $\sigma_v$ ($p<0.001$),  $M_{X_{CO,B}}$ ($p<0.001$), and excitation temperatures $T_{ex}$ ($p<0.001$), and lower virial parameters $\alpha_{\rm{vir}}$ ($p=0.005$, see \S\ref{subsec:virial}) .  

These findings are consistent with comparisons of YSOs and clumps in other low-Z regions.  In the star-forming complex N159 in the LMC, for example, \citet{nayak_molecular_2018} found that CO clumps containing YSOs were more massive than clumps that did not contain YSOs, and that massive YSOs were typically associated with the most massive clumps.  Similarly, in the LMC region N55 \citet{naslim_alma_2018} observed that molecular cores containing YSOs possessed larger linewidths and masses than those that did not.

The majority of clumps (83 out of 110, or $\sim$75$\%$) are not associated with any YSOs, but may have been in the past.  We calculated the distance between each clump and its nearest YSO and found moderate negative correlations between distance to the nearest YSO and clump mass ($\rho_s=-0.48$, $p<0.001$), and distance and $\sigma_{v}$ ($\rho_s=-0.47$, $p<0.001$).  We estimated the distance that a clump could have plausibly travelled since formation of this generation of YSOs began 1 -- 2 Myr ago (C11).  The mean $\sigma_v$ of all clumps is 0.48 km s$^{-1}$ with a standard deviation of 0.17 km s$^{-1}$.  Using this distribution of $\sigma_v$ as a proxy for the speed at which clumps may be moving relative to each other, we calculated the distance traveled over a timescale of 1.5 Myr at a relative speed of 0.65 km s$^{-1}$ to derive a potential distance traveled estimate of $\sim$1 pc.  Twenty of the 83 clumps that do not contain YSOs have a YSO within this distance.  From Mann-Whitney U-tests, these clumps have significantly larger $\sigma_v$ than clumps outside of this distance ($p=0.002$, mean $\sigma_v$ of 0.52 km s$^{-1}$ vs. 0.41 km s$^{-1}$). 

Only 5 of the 33 YSOs (15\%) are not embedded within the projected 2D boundaries of a clump, so traditional hypothesis testing to compare the properties of YSOs within clumps to YSOs outside of clumps is not appropriate.  A simple comparison of the median masses of the two groups suggests that YSOs inside clumps are more massive than YSOs outside of clumps (median $M=$ 3.4 $M_\odot$ vs 2.3 $M_\odot$) and have higher envelope $\dot{M}$ accretion rates (median $\dot{M}=2.1 \times 10^{-6} \ M_\odot$ yr$^{-1}$ vs $\dot{M}=0.76 \times 10^{-6} \ M_\odot$ yr$^{-1}$).  This suggests that the YSOs not embedded within clumps are currently less actively accreting material, although as discussed in \S\ref{sec:archival} this fitted value is a limited indicator of the actual historical accretion rate of the objects.  

We also checked if CO emission that was not assigned to a clump by the {\tt quickclump} algorithm was present around seemingly isolated YSOs by stacking CO spectra around each YSO within an area equal to three synthesized beams.  We found no robust evidence for strong CO emission around these YSOs (Y198, Y271, Y283, Y290, Y358) with all YSOs having one or fewer $3 \sigma$ detections in $^{12}$CO and $^{13}$CO out of the 105 channels in the stacked spectra.  

\citet{Seale2012} found that a large number of massive YSOs in four LMC GMCs were not associated with any molecular clumps detected with HCO$^+$, which they suggested to be the result of the disruption of clumps on $\lesssim 1$ Myr timescales.  In the 30 Doradus region of the LMC, \citet{nayak_study_2016} found that massive YSOs were more likely to be associated with CO clumps than their low-mass counterparts and concluded that the less-massive YSOs not associated with clumps were likely more evolved than the embedded YSOs, as they would have had sufficient time to dissipate their natal molecular clumps through UV radiation.  We draw similar conclusions that unassociated YSOs may be more evolved than embedded YSOs, and that feedback from YSOs may affecting be the molecular gas on the scale of individual clumps ($\lesssim$1 pc).  We note, though, that small sample sizes involved weaken the power of these conclusions, as do the large uncertainties on the masses and accretion rates of each individual YSO.

\vspace{1cm}

%%%%%%%%%%%%%%%%%%%%%%%%%%%%%%%%%%%%%%%%%%%%%%%%
\subsubsection{PMS Stars}\label{S:pms}

\begin{figure}
    \centering
    \includegraphics[width=0.47\textwidth]{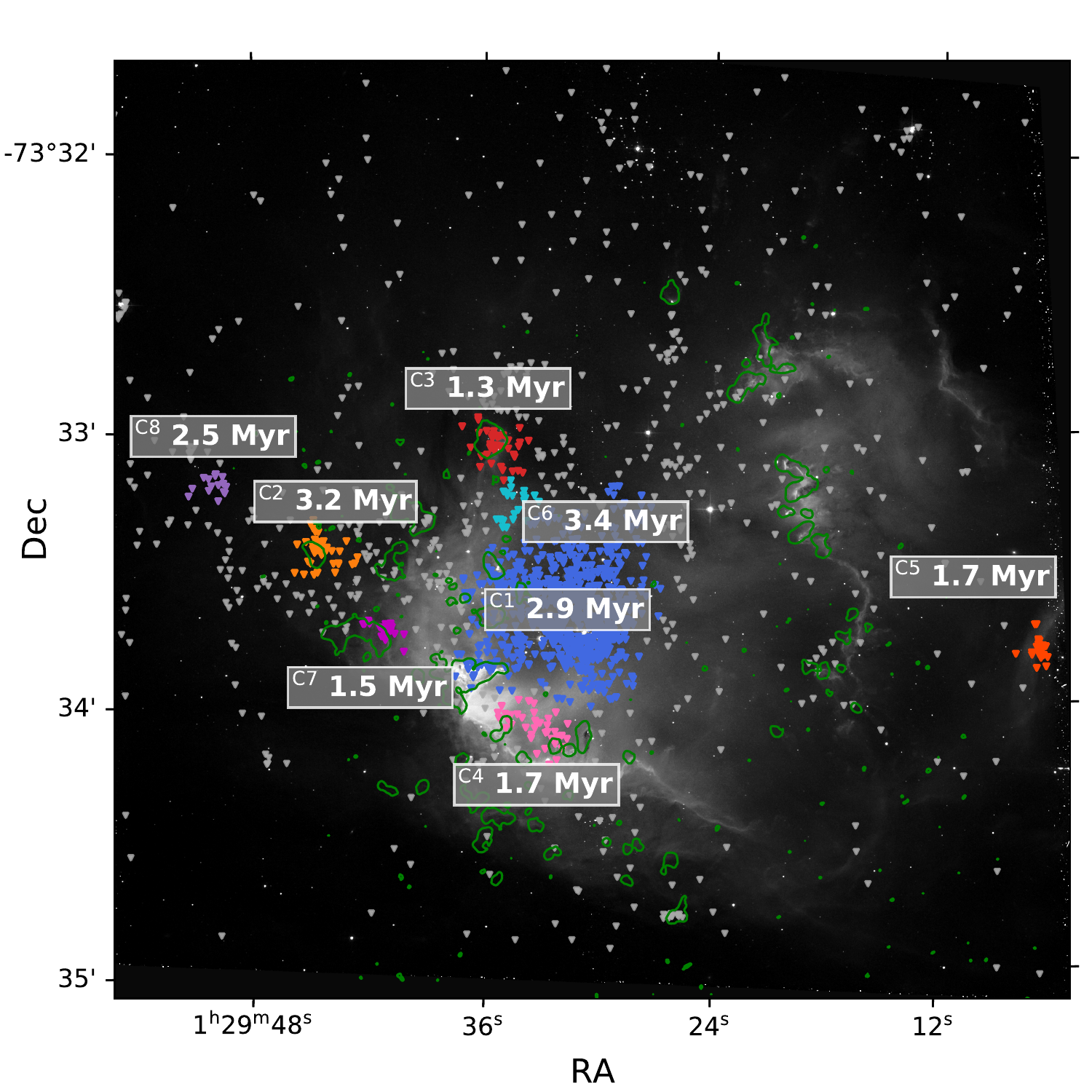}
    \caption{Locations of the PMS star clusters identified by the {\tt DBSCAN} algorithm.  PMS stars are marked with triangles.  The clusters are denoted by color, and are labeled with their Table \ref{tab:PMSclus} ID numbers and the mean ages of their members.  Non-clustered PMS stars are marked with grey triangles.  H$\alpha$ image is in grey and contours of integrated $^{12}$CO(2--1) emission at 0.25 Jy beam$^{-1}$ (km s$^{-1}$)$^{-1}$ is shown in green.}
    \label{fig:pms_dbscan}
\end{figure}

\begin{deluxetable}{cccccc}
\centering
\tablecaption{Properties of PMS Star Clusters \label{tab:PMSclus}} 
\tablehead{  \colhead{ID } &  \colhead{RA} & \colhead{Dec} & \colhead{N$_{\rm{PMS}}$} & \colhead{Mean Age}  & \colhead{$\sigma_{\rm{Age}}$}  \\[-0.1cm] \colhead{}   & \colhead{($\degr$)}  & \colhead{($\degr$)}  & \colhead{}  & \colhead{(Myr)} & \colhead{(Myr)} }
\startdata
 C1 &  22.382 & -73.561 &   779 &  2.9 &      1.6 \\     C2 &  22.436 & -73.557 &    49 &  3.3 &      1.4 \\     C3 &  22.398 & -73.551 &    44 &  1.9 &      1.9 \\     C4 &  22.389 & -73.568 &    41 &  1.8 &      1.5 \\     C5 &  22.278 & -73.563 &    27 &  2.0 &      1.9 \\     C6 &  22.394 & -73.554 &    22 &  3.2 &      1.8 \\     C7 &  22.422 & -73.562 &    21 &  2.0 &      1.9 \\     C8 &  22.459 & -73.553 &    15 &  3.5 &      1.3 \\ 
\enddata 	
\tablecomments{The reported position of each cluster is the mean position of all members of that cluster.   $N_{\rm{PMS}}$ is the number of stars per cluster.}
\end{deluxetable}

G12 analyzed the spatial distribution of PMS stars in N90 and identified 14 sub-clusters.
We supplement this work using the {\tt DBSCAN} algorithm \citep[Density-Based Spatial Clustering of Applications with Noise,][]{ester_1996}, a non-parametric clustering method that defines clusters as regions of high density separated by regions of low density.  It requires the assignment of two parameters: the minimum number of points to form a cluster, MinPts, and the maximum distance over which two points can be considered neighbors, $\epsilon$.  We set the minimum number of cluster members as MinPts = 15 stars.  Following  \citet{rahmah_determination_2016}, we found the optimal value of $\epsilon$ by creating a nearest-neighbors graph of the distance between the k = MinPts nearest-neighbors of points in ascending order and identifying the approximate point of maximum curvature.  This yielded $\epsilon=4.4$".

We identified 8 distinct clusters of PMS stars, the locations of which are shown in Figure \ref{fig:pms_dbscan} and properties summarized in Table \ref{tab:PMSclus}.  Like G12, we find overdensities of PMS stars in the central NGC 602 association and along the rim of the \HII{} region.  The largest cluster, Cluster 1, is centered around NGC 602 and consists of 779 members ($\sim$50$\%$ of the total PMS sample).  \citet{oskinova_discovery_2013} analyzed \textit{Chandra} and \textit{XMM-Newton} observations of N90 and found evidence for extended X-ray emission around this central cluster, and also in a feature on the rim directly to the north of NGC 602 where the {\tt DBSCAN} algorithm identifies Cluster $\#3$.  They attributed these features to the effects of many low-mass PMS stars and YSOs unresolved in the X-ray data, which is consistent with the high concentration of resolved PMS stars in these locations that we find here. 

Through the isochrone fitting described in \S\ref{sec:archival} and shown in Figure \ref{fig:pms_iso}, we find that the majority of the PMS stars in N90 are consistent with ages less than 5 Myr, with a mean age of 3 Myr.  We note that the uncertainties on these estimates are large due to the simple method of isochrone matching that we adopted.  Cluster 1 has a mean age of $\sim$3 Myr, which is consistent with the age estimated for PMS stars in NGC 602 by C11 and G12.  Clusters 3, 4, and 7 are scattered along the \HII{} rim and possess slightly lower mean ages of $\sim$2 Myr.  This would seemingly support a star formation history in which the formation of these clusters was triggered by the formation of Cluster 1.  

However, in all clusters the dispersions in member age are large (typically $\gtrsim$1.5 Myr).  Additionally, there is a notable spread across V--I in the entire sample of PMS stars, with a handful of extremely young, red stars appearing to exist (m$_{555}$ - m$_{814}$ $\gtrsim 2.5$ mag).  Visual inspection reveals that many of the reddest sources are located in extremely crowded areas (for example, Cluster 5 at the edge of the HST coverage).  These sources could be genuine, or could simply be a result of confusion or indicate the presence of misclassified asymptotic giant branch stars or unresolved background galaxies.  

Many of the reddest sources are also associated with bright CO emission along the rim; this suggests that significant differential reddening could be resulting from the brightest clumps.  Of the 1569 PMS stars, 130 fall within the projected boundaries of a CO clump.  These sources appear significantly younger ($p < .001$ from a one-sided Mann-Whitney U test, median age of 2 vs 3.5 Myr), i.e. redder, than PMS stars not inside CO clumps. Conversely, 40 of the 110 CO clumps contain at least one PMS star. Clumps that contain PMS stars have significantly higher masses ($p < 0.001$), surface densities ($p < 0.001$), and $\sigma_v$ ($p = 0.003$) than clumps that do not contain PMS stars, and significantly lower virial parameters $\alpha_{\rm{vir}}$ ($p<0.001$, see \S\ref{subsec:virial}).

Clusters 3 and 7 along the H$\alpha$ rim (with apparent mean ages of 1.9 and 2.0 Myr, respectively) are both associated with strong CO emission.  \citet{de_marchi_photometric_2013} observed that the youngest and reddest PMS stars in N90 were located in dense areas of the rim, and suggested that if the extinction towards these stars was significantly higher than the rest of the sample ($A_V \simeq 2.25$ mag vs. $A_V = 0.25$ mag) the ages of the outlier stars would be comparable to the ages of the PMS stars in the central NGC 602.  In the N83 region of the SMC Wing, \citet{lee_relationship_2015} derived a relationship between CO intensity and extinction, $I_{CO}$/$A_V$ = 1.5 K km s$^{-1}$ (mag)$^{-1}$.  The mean $I_{CO}$ within clumps in N90 that contain PMS stars is 4.7 K km s$^{-1}$, which corresponds to $A_V \sim 3.1$ mag; a centrally embedded star might experience half of this value, $A_V \sim 1.6$ mag.  Thus, reddening resulting from the CO clumps would be sufficient to cause Clusters 3 and 7 to appear much younger than they truly are.  If this is the case, we would then find no strong evidence for a triggered star formation scenario.  

Additionally, we find no correlation between age and radial distance from the center of NGC 602 in the full sample of PMS stars ($\rho_s = 0.08$, $p < 0.001$).  The crossing time through the $\sim$30 pc \HII{} region is 3 Myr, which is equal to the mean ages of the entire PMS sample and most contained clusters.  Since many of the PMS stars are located on the edges of the \HII{} region and the spatial distribution of their ages is roughly uniform, this suggests a sequential star formation scenario is more likely than a triggered event (i.e., the formation of local subclusters of PMS or OB stars would not have had time to directly influence the formation of other subclusters).

%%%%%%%%%%%%%%%%%%%%%%%%%%%%%%%%%%%%%%%%%%%%%%%%%%%%%%%%%%%%%%%%%%%%%%%%%%%%%%%%%%%%%%%%%%%%%%%%%%%%%%%%%%%%%%%
\section{Molecular Gas \& Star Formation in Low Metallcity Environments}\label{sec:gasstars}

%%%%%%%%%%%%%%%%%%%%%%%%%%%%%%%%%%%%%%%%%%%%%%%%%%%%%%%%%
\subsection{Size-Linewidth-Surface Density Relationships}\label{subsec:sizeline}
\begin{figure*}
    \centering
    \includegraphics[width=\textwidth]{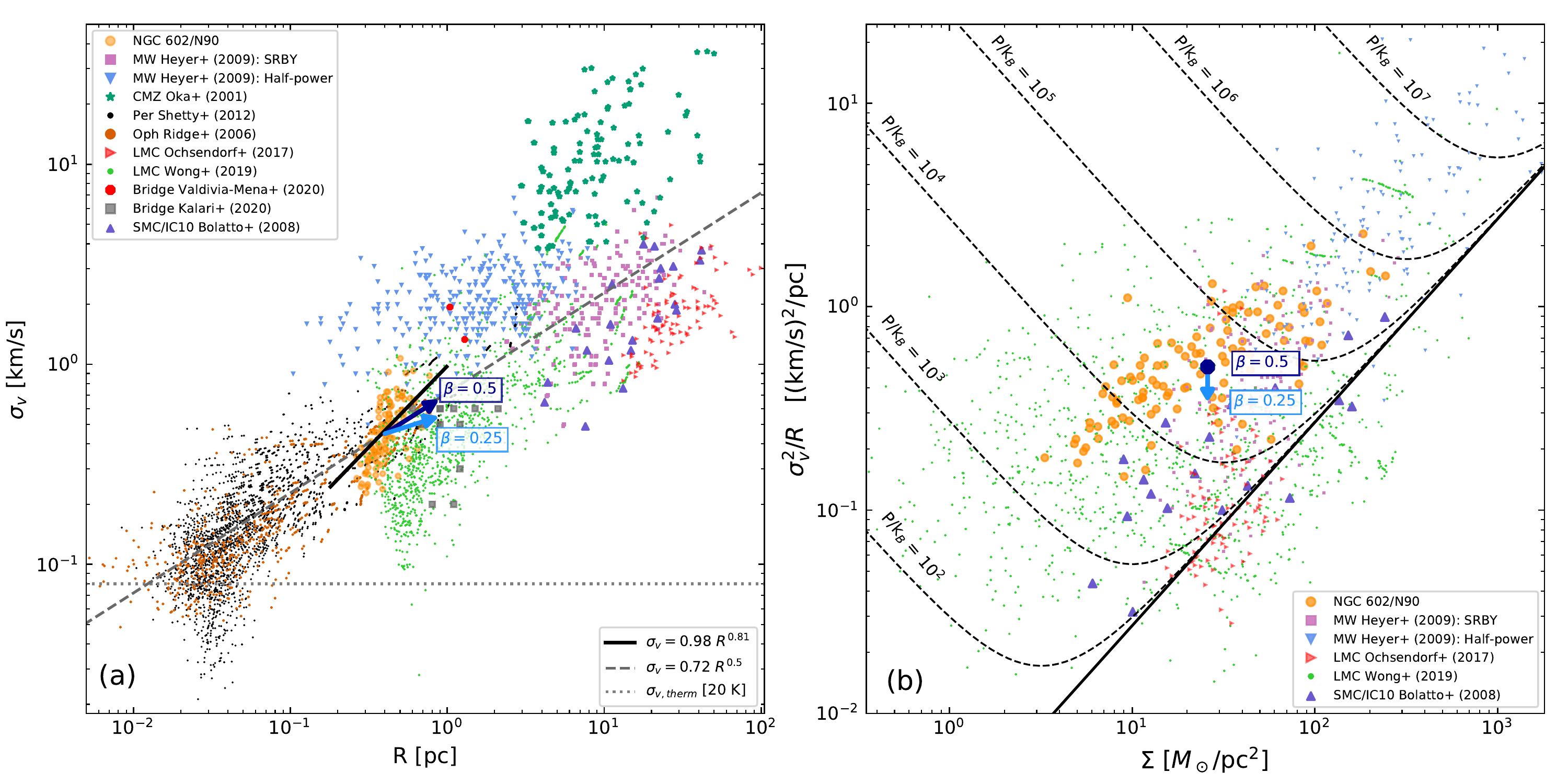}
    \caption{Size-linewidth-surface density plots, adapted from \citet{ONeill2022_codark} with the addition of N90 clumps.  \textit{(a:)} Velocity dispersion $\sigma_v$ compared to radius $R$ of clumps from studies described in \S\ref{subsec:sizeline}.  The arrows show the corrections needed to account for CO-dark gas on observable CO clump properties for a typical clump with [$R_{CO}$=0.40 pc, $M(R_{CO})=13 \ M_\odot$, $\sigma_v(R_{CO})=0.45$ km s$^{-1}$].  The arrows move from the properties of the observed CO clump to the inferred full state of the clump including CO-dark gas.  The dark blue arrow follows a velocity profile $\sigma_v \propto R^{0.5}$ [$\beta=0.5$] and the light blue arrow follows a velocity profile $\sigma_v \propto R^{0.25}$ [$\beta=0.25$].  Both arrows show the effect for a density profile $\rho \propto R^{-1}$ [$k=1$] and $f_{DG}\sim 0.8$.  The grey dashed line follows the relationship $\sigma_{v}=0.72R^{0.5}$ derived by SRBY and the grey dotted line is the expected $\sigma_v$ at T=20 K from thermal motion.  The black solid line was fit to the N90 clumps and follows $\sigma_{v}=0.98 \ R^{0.81}$.
    %%%%%%%%%%%%%%
    \textit{(b:)} Size-linewidth parameter $\sigma_v^2 / R$ compared to surface density $\Sigma$, with correction arrows as in (a).  The black line corresponds to virial equilibrium (Equation \ref{eqn:virial_sigma} with $k=0$), and the black curves correspond to virial equilibrium under varying degrees of external pressure (with units of P/k in K cm$^{-3}$, Equation \ref{eqn:field_extP} with $\Pi=0.6$). }
    \label{fig:RadSigvSurfp}
\end{figure*}

The relationship between size, linewidth, and surface density in molecular clouds has been extensively studied as a proxy for their dynamical states.  \citet{larson1981} identified correlations between global cloud properties including size $R$, linewidth $\sigma_v$, and surface density $\Sigma$.  The first of these correlations follows
\begin{equation} 
    \sigma_v = C \left(\frac{R}{1 \ \rm{pc}}\right)^{\Gamma} \rm{km \ s}^{-1},
    \label{eqn:h09_sigR}
\end{equation}
with $\Gamma=0.5$ and $C \simeq 0.72$ km s$^{-1}$ pc$^{-0.5}$ \citep[e.g.,][hereafter SRBY and H09, respectively]{solomon_mass_1987,heyer_re-examining_2009}.  A virialized spherical clump described by a power-law density distribution $\rho \propto r^{-k}$ should additionally follow the relationship
\begin{equation}
    \frac{\sigma_v^2}{R} = \frac{(3-k)}{3(5-2k)}  \pi G \Sigma.
    \label{eqn:virial_sigma}
\end{equation}

Figure \ref{fig:RadSigvSurfp} compares $\sigma_v$, $R$, and $\Sigma$ for the N90 clumps.  We also show cores, clumps, and GMCs from:
\begin{itemize} \itemsep -0.3 em
    \item Two samples of Galactic GMCs observed by H09, with the first defined from SRBY's $^{12}$CO GMCs and the second from the half-power contours of the central GMCs cores observed in $^{13}$CO (median radii $R$ of 9.7 pc and 1.6 pc, respectively)
    
    \item clouds in the Galactic center studied by \citet{oka_statistical_2001} in \twelveCO  (median $R$ of 8.7 pc)
    
    \item clouds in the Ophiuchus molecular cloud in \thirtCO from \citet{ridge_complete_2006} (median $R$ of 0.07 pc)
    
    \item cores identified using the dendrogram algorithim within the Galactic molecular cloud Perseus A observed in \thirtCO  by \citet{shetty_linewidth-size_2012} (median $R$ of 0.07 pc)
    
    \item clumps in the Magellanic Bridge studied by \citet{kalari_resolved_2020} in CO (1--0) (mean $R$ of 1.1 pc)
    
    \item clumps in the Magellanic Bridge studied in $^{12}$CO (2--1)  by \citet{valdivia-mena_alma_2020} (mean $R$ of 1.1 pc)
    
    \item GMCs identified by \citep{ochsendorf_what_2017} in \twelveCO in $\sim$150 LMC star-forming regions (median $R$ of 27 pc)
    
    \item Clumps identified by \citet{wong_relations_2019} in \twelveCO in the LMC regions 30 Doradus, A439, GMC 104, GMC 1, PCC, and N59C (median $R$ of 1 pc)
    
    \item GMCs in the SMC and dwarf galaxy IC 10 observed in CO (2--1) and (1--0) by \citep{bolatto2008} (median $R$ of 15 pc)

\end{itemize}

Figure \ref{fig:RadSigvSurfp}a compares $\sigma_v$ to $R$. The expected contribution to linewidth by thermal motion at T=20 K, $\sigma_{v,\rm{th}}\sim 0.08$ km s$^{-1}$, is also shown.  We fit a relationship for the N90 clumps of $\Gamma=0.81 \pm 0.10$ and $C=0.98 \pm 0.09$ km s$^{-1}$.  Similarly steep values of $\Gamma$ compared to SRBY have been derived in other low-Z, low-density regions throughout the SMC, LMC, and other local dwarf galaxies (with $\Gamma \sim$ 0.55 -- 0.85 and $C \sim$ 0.2 -- 0.6 km s$^{-1}$, \citealt{bolatto2008, hughes2010_magma, Hughes2013, wong_relations_2019, FinnIndebetouw2022}).  Our fit $C$ is significantly higher than derived in these studies, suggesting that the clumps in N90 have larger linewidths at a given size than structures in those other dwarf galaxy regions.     

\citet{KepleyLeroy2016} studied 8 GMCs in the low-Z (Z $ \sim $ Z$_{SMC}$) starburst dwarf galaxy $\scriptstyle\mathrm{II}$ Zw 40 and derived similarly high $C$; they attributed this to high linewidths and surface densities stemming from a merger between dwarf galaxies that triggered the starburst, rather than high external pressure supporting the GMCs against collapse.  \citet{ImaraFaesi2019} identified $\sim$120 GMCs in the moderately low-Z ($Z \sim 0.7 Z_{\odot}$) starburst dwarf galaxy He 2-10, and found both higher velocity dispersions, surface densities, and C than in comparable Milky Way clouds.  They also fit a very steep size-linewidth slope of $\Gamma$ = 1.3, which they suggested could be the result of tidal interactions or energy and momentum injected from nearby superstar clusters.  

Figure \ref{fig:RadSigvSurfp}b compares surface density $\Sigma$ to the size-linewidth parameter $[\sigma_v^2/R]$.  Virialized clumps are expected to follow Equation \ref{eqn:virial_sigma}, but the N90 clumps have higher $\sigma_{v}^2/R$'s for a given $\Sigma$ than would be expected based on from this trend.  A likely reason that clumps in a region would deviate from expected trends in size-linewidth space, and have higher kinetic energy at a given size scale, is the OB stars in the center of NGC 602 injecting kinetic energy into the surrounding gas. 

If there is significant inter-cloud thermal pressure acting on the molecular clumps, as observed by \citet{oka_statistical_2001} in the Galactic center, and the clumps are assumed to be in virial equilibrium with that pressure, turbulence can be treated as a pressure term \citep{field_does_2011}. This increase in internal turbulent pressure would then be reflected as higher linewidths than expected, as observed in the relationship between $\sigma_v^2 / R$ and $\Sigma$ in clumps in N90.  The external pressure $P_e$ for a clump to remain bound (under the assumption that they are virialized, which is discussed in \S\ref{subsec:virial}) can be found using \citep{elmegreen_pressure_1989},
\begin{equation}
    \label{eqn:extP_elmegreen}
    P_e = \frac{3 \Pi M \sigma_v^2}{4 \pi R^3}.
\end{equation}
The median external pressure required for clumps to remain bound is $P_e/k_B \sim (2.4 \pm 1.3) \times 10^4$ K cm$^{-3}$. Clumps on the rim require on average 1.5$\times$ larger external pressures to remain bound than clumps that are not on the rim, but a K-S test reveals no overall .significant difference between the distributions of the two groups ($p = 0.14$).  

The surface density $\Sigma$ and size-linewidth parameter $\sigma_v^2 / R$ for clumps in pressure-bounded virial equilibrium can be related to $P_e$ as \citep{field_does_2011}, \begin{equation}
    \frac{\sigma^2}{R} = \frac{1}{3} \left ( \pi \Pi G \Sigma + \frac{4 P_{e}}{\Sigma} \right).
    \label{eqn:field_extP}
\end{equation}
We adopt $\Pi = 0.6$ for a uniform sphere \citep{field_does_2011}.  These relationships are represented in Figure \ref{fig:RadSigvSurfp}b by the black V-shaped curves for $P_e/k_B$ between 10$^{2}$--10$^{7}$ K cm$^{-3}$.  The N90 clumps do not appear consistent with being supported by any one value of $P_e$, but the majority of the clumps require between $P_e = 10^3$ K cm$^{-3}$ and $10^6$ K cm$^{-3}$.

The clumps identified by \citet{wong_relations_2019} in six GMCs in the LMC with $^{12}$CO and $^{13}$CO emission were observed at comparable angular resolution to our data (having a synthetic beam size of 3.5" [0.8 pc], vs 1.3" [0.4 pc] in the data we present here).  They found higher $\sigma_v$ at a given $R$ in regions with higher infrared surface brightness (a plausible measure of star formation rate and stellar feedback).  The mean 8$\mu$m brightness of $\sim$0.5 MJy/sr in N90 corresponds to the lower end of surface brightnesses in the \citet{wong_relations_2019} study of LMC regions.  Thus N90 is consistent with the \citet{wong_relations_2019} findings in this parameter space, although care should be exercised since both measures may be affected by reduced metallicity. The 8$\mu$m diffuse emission at a given radiation intensity is lower in the SMC than LMC, and the CO measurements may also be affected as discussed below. 

\vspace{1cm}

%%%%%%%%%%%%%%%%%%%%%%%%%%%%%%%%%%%%%%%%%%%%%%%%%%%%%%%%%%%%%%%%%%%% 

\subsection{``CO-dark'' Gas: Effects of Low Metallicity on CO Diagnostics in N90}\label{subsec:dark}

At low metallicities, reduced dust-to-gas ratios and typically stronger radiation fields increase the efficiency of CO destruction \citep{madden_ism_2006, gordon_surveying_2011, madden2020} and decrease the effectiveness of CO as a tracer for H$_2$.  The C$^{+}$/C$^{0}$/CO transition retreats further into the center of clumps, which causes the fraction of ``CO-dark'' H$_2$ gas mass not traced by CO to increase \citep{wolfire_dark_2010, glover_relationship_2011}.  It is possible that the discrepancies in N90 clump properties compared to expected size-linewidth-surface density trends are the result of this increased proportion of CO-dark gas.  We apply corrections derived by \citet[][hereafter O22]{ONeill2022_codark} to account for the expected contribution of CO-dark gas on observed clump properties in N90.  

As defined by \citet{wolfire_dark_2010}, the fraction of H$_2$ gas mass that is CO-dark in a clump, $f_{DG}$, can be expressed as
\begin{equation}
    f_{DG} = 1 - \frac{M(R_{CO})}{M(R_{H_2})}, 
    \label{eqn:fdg_Gen}
\end{equation}
where $M(r)$ is the mass contained within a given radius $r$, $R_{CO}$ is the radius of the CO-traceable material, and $R_{H_2}$ is the radius at which half of the hydrogen in the clump's diffuse envelope is molecular and half is atomic.  Under typical Galactic conditions, $f_{DG}$ is found to be $\gtrsim$0.3 \citep[e.g.,][]{grenier2005,Abdo2010,velusamy2010,lee2012,langer2014,xu2016}.

To apply the corrections derived by O22, we assume that clumps follow a power-law density profile with $k=1$ (Equation \ref{eqn:density_prof}) 
and internal velocity dispersion profile, $\sigma_{v}(r)$, of 
\begin{equation}
    \sigma_{v}(r) = \left(\frac{r}{R_{CO}}\right)^{\beta} \sigma_{v}(R_{CO}).
\end{equation} 
$\beta \sim$0.2--0.3 is common in observations and simulations of young cores and clumps \citep{caselli_1995,Tatematsu2004,lee2015_sfr,lin2021}, while a steeper $\beta\sim$0.5 in alignment with the global size-linewidth relationship of SRBY with $\Gamma=0.5$ (Equation \ref{eqn:h09_sigR}) has been found to hold on larger scales \citep[e.g.,][]{HeyerBrunt2004, Dobbs2015}.  Based on these findings, we adopt $\beta = 0.25$ for the bulk of our analysis, but also present results for $\beta = 0.5$. 

We estimate an appropriate $f_{DG}$ for N90 based on measurements of CO-dark gas content in other low-Z, low-density environments.  In the LMC, \citet{chevance_co-dark_2020} derived $f_{DG}$ $\gtrsim$0.75 in the star-forming region 30 Doradus, and in the $\textrm{H} \scriptstyle\mathrm{II}$ region N11, \citet{Lebouteiller2019_n11} found that the majority (40--100\%) of molecular gas was CO-dark.  Throughout the low-Z outskirts of the Milky Way \citet{pineda2013} derived $f_{DG}\sim$0.8, while in the nearby low-metallicity dwarf galaxy NGC 4214 ($Z\sim$0.3--0.4 $Z_\odot$, \citealt{Hermelo2013_ngc4214}), \citet{Fahrion2017_ngc4214} derived $f_{DG}=0.79$.  \citet{Pineda2017_smc} found that 77\% of the total molecular gas in their sample of 18 line-of-sight pointings across the SMC was CO-dark H$_2$,  and in the SMC regions N66, N25+N26, and N88 (located in the northern Bar, southwest Bar, and Wing, respectively). \citet{RequenaTorres2016_carbonSMC} derived a typical fractional abundance of CO-dark gas to be 80--95\%.  In four star-forming regions in the nearby southwest Bar of the SMC with cloud density is comparable to N90 and the Wing ($\bar{N} \lesssim 2 \times 10^{21}$ cm$^{-2}$), \citet{jameson_first_2018} derived an average $f_{DG} \simeq 0.8$.  Based on these results, we adopt $f_{DG}\sim 0.8 \pm 0.1$ for an estimated dark-gas fraction in N90 for the remainder of this work. 
   
O22 derived corrections for clump properties under a power-law density profile, including
\begin{equation}
\begin{split}
    R_{H_2} & =   [1-f_{DG}]^{1/(k-3)} \ R_{CO}, \\
    \sigma_v(R_{H_2}) & = [1-f_{DG}]^{\beta / (k-3)} \ \sigma_v(R_{CO}), \\ 
    M(R_{H_2}) & =  [1-f_{DG}]^{-1} \ M(R_{CO}), \\ 
    \Sigma_{H_2} & =  [1-f_{DG}]^{(1-k)/(k-3)} \ \Sigma_{CO}.
\end{split}
\label{eqn:CDG_cors}
\end{equation}
After applying these corrections to the $M_{X_{CO,B}}$-derived clump masses, which estimate the mass within the CO-traced regions, we find a total molecular gas mass in N90 of $M_{DG} \sim16,600  \pm 2,400  \ M_\odot$. The errors on this estimate are derived from the combined uncertainties of $f_{DG}$ and $M_{X_{CO,B}}$. 

In Figure \ref{fig:RadSigvSurfp}a and Figure \ref{fig:RadSigvSurfp}b, we demonstrate the effects of these corrections on a typical observed clump with properties [$R_{CO}$=0.40 pc, $M(R_{CO})=13 \ M_\odot$, $\sigma_v(R_{CO})=0.45$ km s$^{-1}$, $\Sigma_{CO} = 26 \ M_\odot$ pc$^{-2}$].  The arrows show the change in placement in size-linewidth-surface density space from the example clump's observed CO properties to its inferred ``true'' characteristics when including CO-dark gas for $\beta=0.25$ and $\beta=0.5$.  Assuming $f_{DG} = 0.8$ and $k=1$, the preferred $\beta=0.25$ yields corrected properties of [$R_{H_2} = 0.9$ pc, $M(R_{H_2}) = 65 \ M_\odot$, $\sigma_{v,H_2}=0.55$ km s$^{-1}$, $\Sigma_{H_2} = 26 \ M_\odot$ pc$^{-2}$].  In this case, the corrections bring clump properties closer to agreement with expected $\Sigma$--$[\sigma_v^2/R]$ trends.  Since this change is only by a relatively small amount, though, and O22 demonstrated that the effects of these corrections vary significantly depending on the density and velocity profiles assumed, we conclude that is unlikely but not impossible that enhanced photodissociation in this low-Z environment is responsible for observed departures from size-linewidth-surface density trends in N90.

\subsection{Stability of Clumps}\label{subsec:virial}

\begin{figure}
    \centering
    \includegraphics[width=0.47\textwidth]{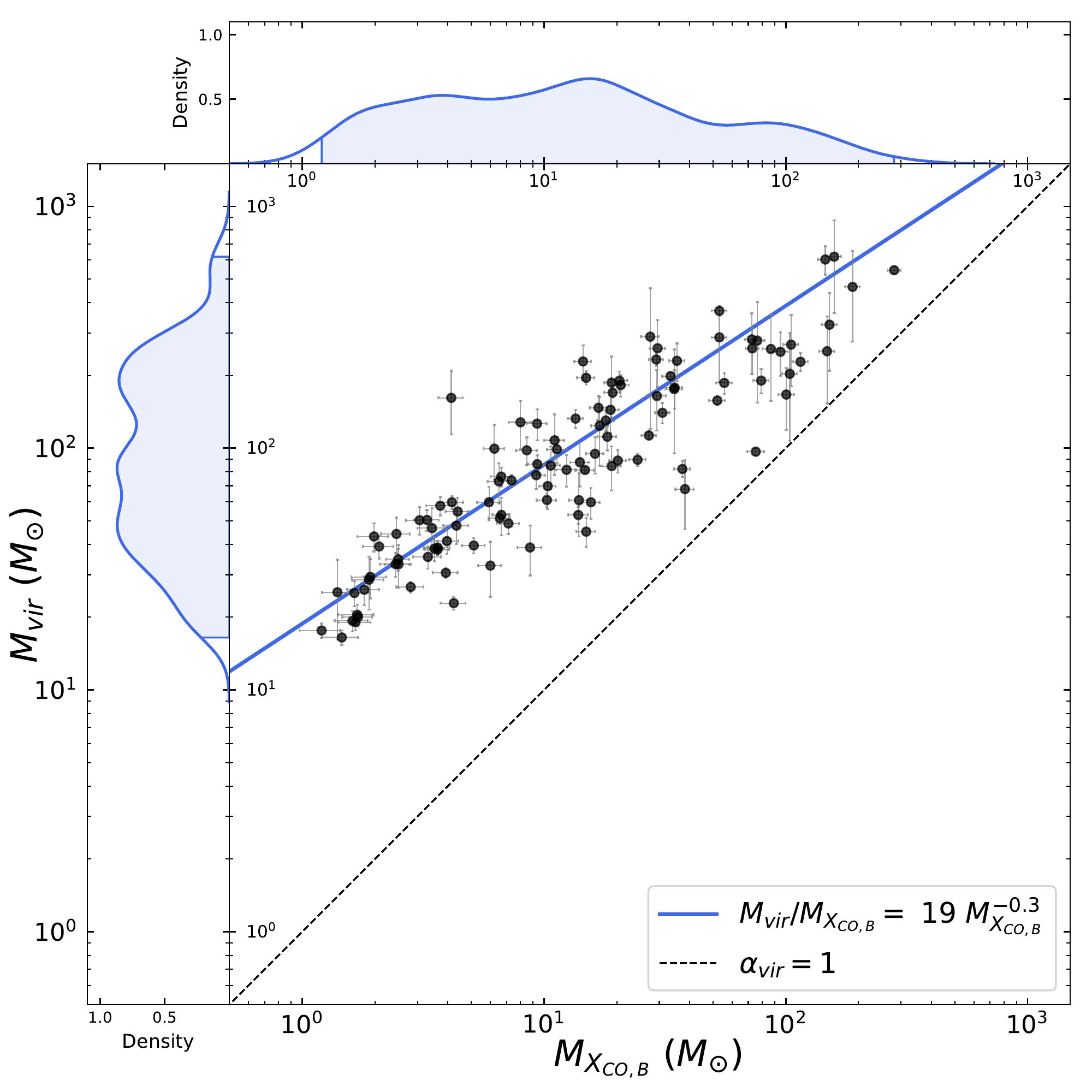}
    \caption{Clump virial mass $M_{\rm{vir}}$ is compared to clump $X_{CO,B}$-derived mass $M_{X_{CO,B}}$.  The black dashed line shows a 1:1 relationship where $M_{\rm{vir}}$ = $M_{X_{CO,B}}$, i.e., where the virial parameter $\alpha_{\rm{vir}}=1$ and stability could be expected, and the blue line shows a best-fit relationship $M_{\rm{vir}}$/$M_{X_{CO,B}}$ = 19 $M_{X_{CO,B}}^{-0.34}$.  The two mass distributions are represented individually by Gaussian kernel density estimations placed horizontally on the top axis of the scatter plot ($M_{X_{CO,B}}$) and vertically on its left axis ($M_{\rm{vir}}$).
    }
    \label{fig:Virial_DG_scatter}
\end{figure}

The evolution and stability of clumps can be also studied through assessing their virial masses, which for a clump with a power-law density profile (Equation \ref{eqn:density_prof}) can be found as \citep{solomon_mass_1987,MacLaren1988}
\begin{equation}
    M_{\rm{vir}}(r) = \frac{3(5-2 k)}{(3-k)} \  \frac{r \ \sigma_{v}^2(r)}{G},
\end{equation} 
where $k$ is the index of the power-law.  As in \S\ref{subsec:dark} we assume $k = 1$.  Virial masses for clumps in N90 ranged from 16 $M_{\odot}$ to 620 $M_{\odot}$, with a median  mass of 85 $M_{\odot}$ and total virial mass of all clumps of 14,045 $\pm$ 515 $M_{\odot}$.  

We compare the relationship between individual $M_{\rm{vir}}$ and $M_{X_{CO,B}}$ in Figure \ref{fig:Virial_DG_scatter}.  A relationship $M_{\rm{vir}}$/$M$ $\propto$ $M^{-\eta}$ has been observed to hold for dense clumps in many regions, with estimates of $\eta$ typically ranging from $\eta \sim 0.3$ to $0.4$ in Galactic clouds \citep[e.g.,][]{Yonekura1997,ikeda2009}.  A least-squares fit for clumps in N90 yields $M_{\rm{vir}}$/$M_{X_{CO,B}} = (18.8 \pm 1.3) \ M_{LTE}^{(-0.34 \ \pm \ 0.04)}$.  KDEs of the distributions of $M_{\rm{vir}}$ and $M_{X_{CO,B}}$ are also shown, with $M_{\rm{vir}}$ being centered at higher masses than $M_{X_{CO,B}}$.  

We then calculated the virial parameter $\alpha_{\rm{vir}}$ defined by \citet{bertoldi_pressure-confined_1992} for all clumps as
\begin{equation}
\alpha_{\rm{vir}} = \frac{2 \Omega_K}{| \Omega_G|} = \frac{M_{\rm{vir}}}{M_{X_{CO,B}}},
\label{eqn:alvir}
\end{equation}
where $\Omega_G$ is total gravitational potential energy and $\Omega_K$ is the total kinetic energy.  $\alpha_{\rm{vir}} \sim 1$ indicates that a clump is gravitationally stable, and $\alpha_{\rm{vir}} \gg 1$ indicates that a clump is sub-critical and will likely expand unless confined by external pressure.  There is a wide variation in $\alpha_{\rm{vir}}$ from clump to clump, with $\alpha_{\rm{vir}}$ ranging from 1.3 to 39.  The median value is $\alpha_{\rm{vir}} = 7.95$, with lower and upper quartiles of 4.4 and 11.5, respectively.  

This is significantly higher than many recent measurements of $\alpha_{\rm{vir}} \lesssim 2$ in molecular clouds (see \citealt{kauffmann_low_2013} for a review), and may be an overestimate due to the effects of CO-dark gas, which we discuss below.  In the 30 Doradus region of the LMC, \citet{wong_alma_2017} found that virial clump masses were typically an order of magnitude larger than CO-derived masses, which is similar to what we observe here.  \citet{schruba_physical_2017} studied $\sim$150 small CO clumps (mean radius $R \simeq 2.3$ pc) in five star-forming regions in the $Z = 1/5 \ Z_\odot$ dwarf galaxy NGC 6822 and found large values of $\alpha_{vir}$ (from $\sim$1 to $\gtrsim$10).  Similarly high values of $\alpha_{vir}$ have also been found in clumps in the Galactic Central Molecular Zone \citep{MyersHatchfield2022} and in cores in the Pipe Nebula \citep{lada_nature_2008}.

The increased degree of photodissociation in low-Z environments (see \S\ref{subsec:dark}) compromises the fundamental assumption that CO emission accurately traces clump mass, and by extension interpretations of $\alpha_{\rm{vir}}$ values. We expect the enhanced amount of CO-dark gas in this region to cause $M_{X_{CO,B}}$ to be a significant underestimate of the total amount of molecular gas (\S\ref{subsec:dark}), so correcting for the ``true'' values of $\alpha_{\rm{vir}}$ including CO-dark gas could bring the clumps more in line with expected trends.  O22 derived the CO-dark-corrected virial mass and parameter as
\begin{equation}
\begin{split}
    M_{\rm{vir}}(R_{H_2}) & = [1-f_{DG}]^{(2\beta + 1)/(k-3)} \ M_{\rm{vir}}(R_{CO}) , \\ 
    \alpha_{vir,H_2} & =  [1-f_{DG}]^{(2\beta + k - 2)/(k-3)} \  \alpha_{vir, CO}. 
\end{split}
\label{eqn:fdg_cor_avir}
\end{equation}

We correct clump virial masses and parameters using Equations \ref{eqn:CDG_cors} and \ref{eqn:fdg_cor_avir}, with a  CO-dark gas mass fraction $f_{DG} \sim 0.8$ assumed in \S\ref{subsec:dark} and velocity dispersion profiles following $\beta=0.25$ and $\beta=0.5$.  
For [$k=1$, $\beta=0.5$], values of $\alpha_{vir}$ remain unchanged from the original estimate.   For the preferred [$k=1, \beta=0.25$], the new median value is $\alpha_{\rm{vir}, H_2} = 5.3$.  Although this is a reduction, this is still much higher than would be expected for a virialized clump.  

We note that the calculations of virial masses suffer from large uncertainties stemming from the determination of radii and assumption of a spherical clump, and that the many uncertainties in the calculations of column densities (especially those stemming from assuming equal excitation temperatures for \twelveCO and \thirtCO and constant abundance ratios between $^{12}$CO, $^{13}$CO, and H$_{2}$), $X_{CO,B}$ masses, and the dark-gas mass fraction are also significant.  Still, the only marginal decrease in $\alpha_{\rm{vir}}$ implies that the clumps are either confined by high levels of external pressure or are not evolving near a virialized state.  In any case, it is unlikely that CO-dark gas is responsible for the observed high $\alpha_{\rm{vir}}$.

The high values of $\alpha_{\rm{vir}}$ suggest that the clumps have higher internal kinetic energies than clumps in other regions and galaxies.  If the clumps are long-lived, this imbalance must be addressed by external pressure or magnetic fields.  The data are generally consistent with N90 being an energetic region and contributing a higher inter-clump-medium pressure than in more quiescent regions.  If the central OB cluster is responsible for this energetic state, we would expect to see some variation in virial parameter and other clump properties with proximity to the cluster.

\begin{figure}
    \centering
    \includegraphics[width=0.44\textwidth]{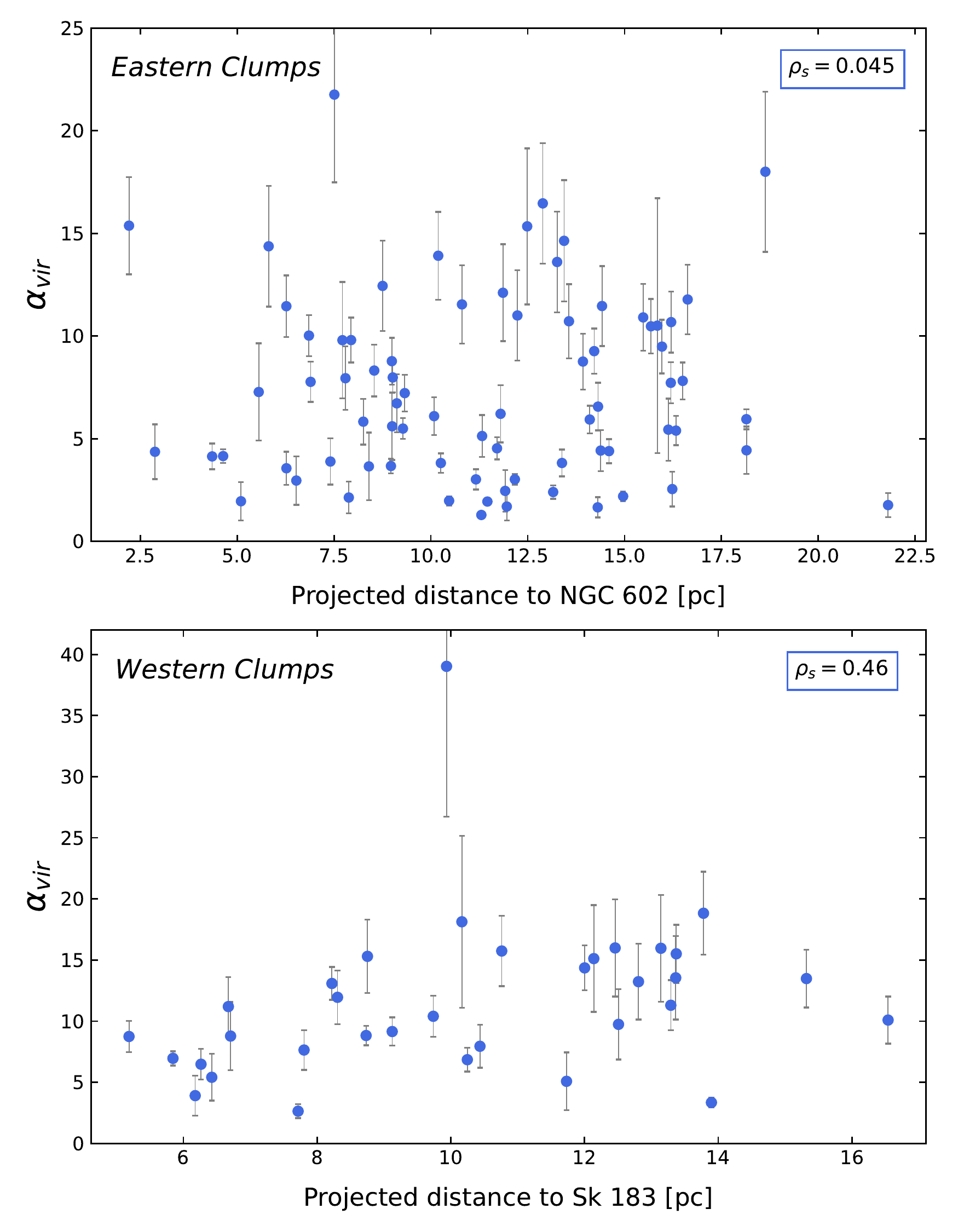}
    \caption{ \textit{Top:} Virial parameter $\alpha_{\rm{vir}}$ is compared to projected distance to NGC 602 for clumps in the Eastern half of N90. The value of Spearman's $\rho_s$ rank correlation coefficient for the two variables is shown in the top right.  \textit{Bottom:} As the top, but for projected distance to the massive star Sk 183 for clumps on the Western half of N90.  }
    \label{fig:ma_avir}
\end{figure}

However, we found no significant difference in the distributions of $\alpha_{\rm{vir}}$ between the rim vs. non-rim clumps ($p=0.095$ from a K-S test).  As shown in Figure \ref{fig:ma_avir}, there is no correlation between $\alpha_{\rm{vir}}$ and distance from NGC 602 for clumps on the Eastern rim ($\rho_s$ = 0.04, $p = 0.7$), while on the Western rim, there is only a moderate trend for increased $\alpha_{\rm{vir}}$ as a function of distance from Sk 183 ($\rho_s$ = 0.46, $p=0.006$).  This is similar to the recent results of \citet{WongOudshoorn2022} and \citet{FinnIndebetouw2022} in the LMC, who both found no significant correlations between clump $\alpha_{\rm{vir}}$ and distance from the super star cluster R136.  The overall absence of any strong trends with position in N90 suggests that the the entire region is energetic for a different reason, e.g., the aftereffects of the collision between the \HI{} clouds/supershells.  

%%%%%%%%%%%%%%%%%%%%%%%%%%%%%%%%%%%%%%%%%%%%%%%%%%%%%%%%%%%%%%%%%%%%%%%%%%%%%%%%%%%%%%
\subsection{Efficiency of Low-Z Star Formation}\label{s:SF}

\begin{figure}
    \centering
    \includegraphics[width=0.47\textwidth]{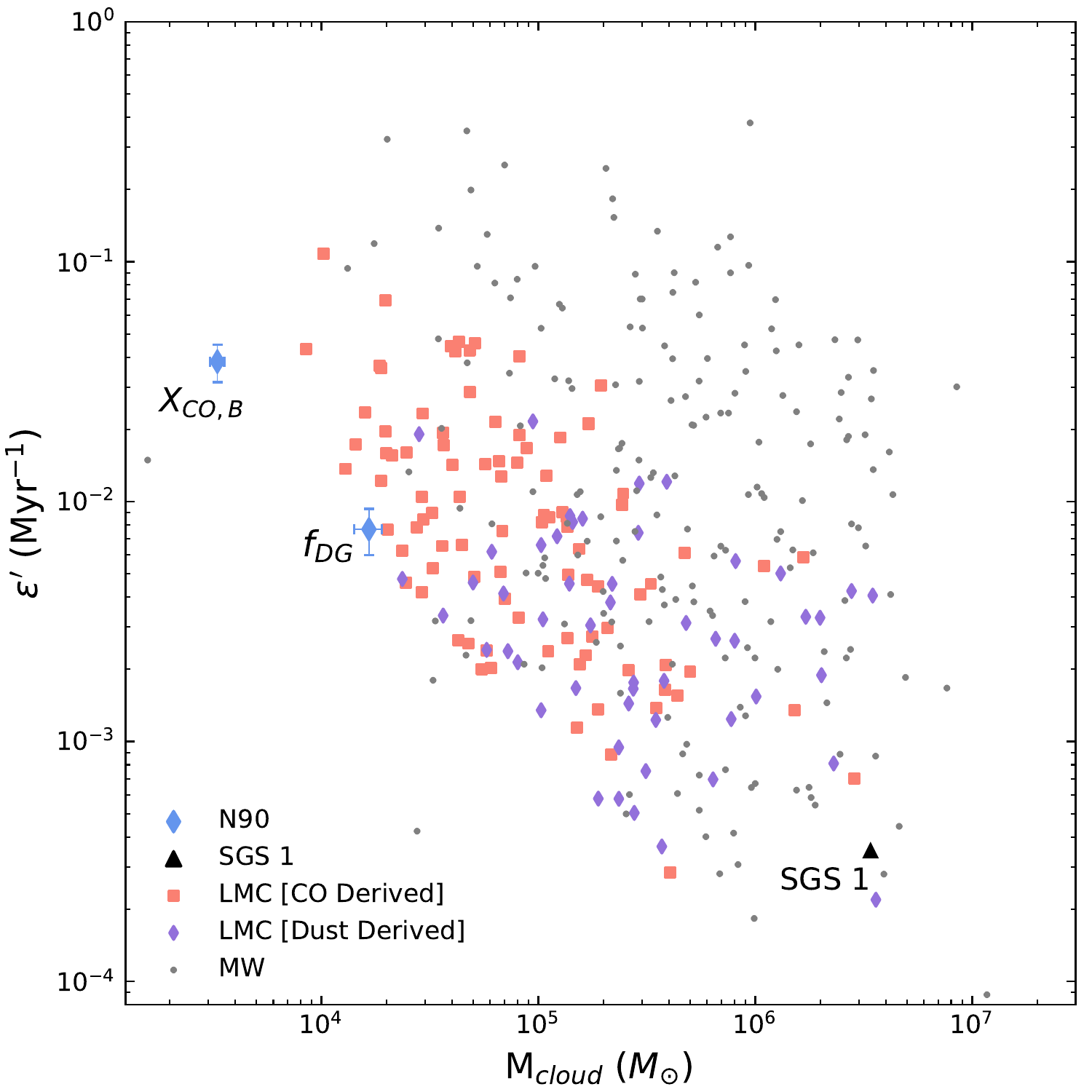}
    \caption{Star formation efficiency $\epsilon$' $=$ SFR/$M_{\rm cloud}$ is compared to cloud mass $M_{\rm cloud}$.  The $\epsilon$' value we derived from the CO-bright gas mass in N90 using an $X_{CO,B}$ conversion factor ($X_{CO,B} = 3.4 \times 10^{20}$ cm$^{-2}$ (K km s$^{-1}$)$^{-1}$ is shown along with the $\epsilon$' derived through correcting for CO-dark gas mass ($f_{DG}\simeq 0.8$, \S\ref{subsec:dark}) by the labeled blue points.  The $\epsilon'$ we derived for SGS 1 is marked with a black diamond.  Results for a sample of LMC GMCs analyzed by \citet{ochsendorf_what_2017} are shown: pink squares had masses derived through $\alpha_{CO}$ conversion, and purple diamonds had masses derived 
    from \citet{Jameson2016}'s dust-based molecular hydrogen map of the LMC.  A sample of Galactic clouds studied by \citet{lee_observational_2016} is also shown by gray circles.}
    \label{fig:epsilonYSO}
\end{figure}

The supergiant shell SGS 1 falls on the boundary between the \HI{}-dominated outskirts and molecule-rich center of the SMC, and the star formation rate (SFR) per unit area and mass has been observed to drop dramatically along such transitions \citep{krumholz_star_2013}.  The low amount of CO emission to the south of N90 compared to the concentrated CO emission to its north may reflect this transition in the Wing.  When gas mass is assessed using CO emission, low-Z environments have been observed to have higher star formation rates (SFRs), and by extension higher apparent star formation efficiencies (SFEs) than higher-Z regions \citep[e.g.,][]{Galametz2009,Schruba2012,schruba_physical_2017}, although this may simply be the result of CO being a poor tracer of H$_2$ at low Z.  Here we explore metrics of SFE in N90.

We initially examined the SFE in N90 at the scale of individual clumps ($\lesssim$1 pc).  If one calculates a by-clump efficiency as the ratio of the mass of any contained YSOs to CO-dark corrected clump mass, efficiency appears to decrease with clump mass, with values ranging from 0.5$\%$ to $16.5\%$; however, this ratio cannot account for the unknown difference between the original clump mass that created a given YSO and the observable extant gas mass.  Since the gas within a few parsecs of a YSO is quickly disrupted by the star formation process and the dynamic range of the Spitzer-identified YSO masses is small, it is difficult to conclude much from such a clump-scale efficiency.

Although the offset in time between current and original gas mass is an unavoidable limitation of SFE measurements, comparing the total YSO and CO masses is more useful in that it likely averages over a significant portion of the clump evolutionary sequence.  It can then provide a more meaningful efficiency estimate than analysis of individual clumps.  With this in mind, we studied SFE and the SFR throughout the entirety of N90 through analyzing the overall population of candidate YSOs.

To account for incompleteness in the YSO sample, we assumed a two-part stellar IMF of the form dN/d log M $\propto M^{-\alpha}$, where $\alpha = 0.3$ below 0.5 $M_{\odot}$ and $\alpha=1.3$ above 0.5 $M_{\odot}$ \citep{kroupa_variation_2001}. We found no high mass YSOs ($\gtrsim 10 \ M_{\odot}$), which suggests that the IMF for the current generation of star formation in N90 is not fully populated.  We scaled this mass function to the peak of our observed YSO mass distribution at 3 $M_{\odot}$ and integrated over 0.08 to 50 $M_{\odot}$ to derive a total YSO mass in N90 of $M_{*} \simeq 1250 \ \pm 160 \ M_{\odot}$, which corresponds to an estimated total number of YSOs of N(YSO) $\simeq 260$.  

In comparison, C11 derived a total YSO mass of 2250 $M_{\odot}$ as inferred through the same method.  This difference stems from our globally lower YSO mass estimates.  Our list of YSO candidates is identical to C11's but (all but one of the) revised individual YSO masses are lower than the original estimates, so the peak of our stellar mass function is shifted to lower masses than C11's.  Since C11 and this work both infer the total YSO mass through scaling a \citet{kroupa_variation_2001} IMF to the peak of the YSO mass function, it is unsurprising that we derive a significantly lower total mass estimate.  

Using only the $X_{\textrm{CO,B}}$-derived CO-bright gas mass, this total YSO mass yields a recent SFE ($\epsilon =M_{*}/M_{\rm{gas}}$) in N90 of $\epsilon_{X_{CO,B}} = 38 \pm 7$ \%.  After correcting for CO-dark gas, we estimate an overall $\epsilon_{DG} = (1- f_{DG}) \epsilon_{X_{CO,B}}$ (O22) of $\epsilon_{DG} = 8 \pm 3$ \%.  This estimate is significantly lower than the 20$\%$ formation efficiency estimate that \citet{fukui_formation_2020} derived for N90, which is unsurprising due to our addition of CO-dark gas mass to our calculation. 

We calculated the SFR in N90 as SFR$_{\rm{YSO}}$ = N(YSOs) $\times \bar{M} / t_{*}$ \citep{ochsendorf_what_2017}, where $\bar{M}$ is the mean mass of the fully populated IMF (here 0.5 $M_{\odot}$) and $t_{*}$ is the typical YSO age (here 1 Myr).  We derived a SFR$_{\rm{YSO}} = 130 \pm 30 \ M_{\odot}$ Myr$^{-1}$.  From studying the optical PMS population in N90, \citet{cignoni_star_2009} found that the SFR has been increasing over the last 10 Myr, with a SFR of 150 $M_{\odot}$ Myr$^{-1}$ between 5 and 2.5 Myr ago and reaching a peak of 300 -- 700 $M_{\odot}$ Myr$^{-1}$ in the last 2.5 Myr.  C11 derived a SFR of 2200 $M_{\odot}$ Myr$^{-1}$ over the last 1 Myr through analyzing the YSO MF; we attribute our reduced estimate to our globally lower fitted YSO masses discussed earlier in this subsection.

We then derived the SFE using the notation of \citet{Kennicutt2012} where $\epsilon'$ = SFR / M$_{\rm{gas}}$.  Using the $X_{CO,B}$ CO-bright mass estimate, we estimate that $\epsilon_{X_{CO,B}}' = 0.04 \pm 0.01$ Myr$^{-1}$, while the CO-dark corrected mass estimate yields $\epsilon_{DG}' = 0.01 \pm 0.005$ Myr$^{-1}$.  If the SFRs derived by \citet{cignoni_star_2009} or C11 were used $\epsilon'$ would increase by a small amount, but would not change the conclusions we draw below.

In Figure \ref{fig:epsilonYSO}, these results are compared to values derived for $\sim$150 star forming regions in the LMC through analysis of YSO MFs by \citet{ochsendorf_what_2017} and for a sample of $\sim$190 Galactic clouds analyzed by \citet{lee_observational_2016}.  \citet{ochsendorf_what_2017} and \citet{lee_observational_2016} both found that $\epsilon'$ increased with decreasing cloud masses.  
For comparison, we also derive an estimated $\epsilon'$ for SGS 1.  \citet{rubele_vmc_2018} derived a SFR of $\sim 1.19 \times 10^{-3} \ M_\odot$ yr$^{-1}$ in the last 8 Myr for a 21' $\times$  21.5' region centered in SGS 1.  We assume a typical N(\HI{}) in this area of 2 $\times 10^{21}$ cm$^{-2}$ \citep{welty2012} and a mean atomic mass of $\bar{m} = 1.5 m_{H}$ \citep{nigra_ngc_2008}.  This yields an estimated gas mass within the \citet{rubele_vmc_2018} region of 3.4 $\times$ 10$^{6}$ \ $M_\odot$ and an $\epsilon' = 3.5 \times 10^{-4}$ Myr$^{-1}$, which is significantly lower than the values derived in N90 and consistent with decreasing $\epsilon'$ with increasing gas mass.  

Although on the low end of values observed for similar LMC and Galactic clouds, the SFE we derive for the relatively low-mass N90 is consistent with this trend.  We therefore conclude that star formation in N90 and this region of the Wing of the SMC is not significantly more efficient than in higher-Z environments.

%%%%%%%%%%%%%%%%%%%%%%%%%%%%%%%%%%%%%%%%%%%%%%%
\section{Discussion and Conclusions}\label{sec:conclusions}
%%%%%%%%%%%%%%%%%%%%%%%%%%%%%%%%%%%%%%%%%%%%%%%
\subsection{Evolutionary History of N90}\label{S:evol}

%%%%%%%%%%%%%%%%%%%%%%%%%%%%%%%%% TIMESCALES TABLE
\begin{deluxetable*}{c|c|c}
\centering
\tablecaption{History of N90 \label{tab:timescale}} 
\tablehead{\colhead{Time}	&	\colhead{Event and Derivation Method}	&	\colhead{Reference}}
\startdata	
	&	\textbf{Older Populations Form in SGS 1}	&		\\
$\gtrsim 100$ Myr & SF begins in SGS 1, from SED and HRD fitting & \citet{ramachandran_testing_2019} \\
 $\sim$50 Myr	&	Subclusters NGC 602 B and NGC 602 B2 form, from optical CMD fitting	& \citet{de_marchi_photometric_2013}	\\	
 25 -- 40 Myr & Extended SF event in SGS 1, from optical/near-UV CMD fitting & \citet{fulmer_testing_2020}  \\ 
 \hline 
%%%%%%%%%%%%%%%%%%%%%%%%%%%%%
	&	\textbf{Formation event for NGC 602} 	&		\\
8 Myr	&	Collision between clouds, from separation in \HI{} velocity components	&	\citet{fukui_formation_2020}	\\
7 Myr	&	Collision between shells, from expansion velocities of \HI{} shells	&	\citet{nigra_ngc_2008}	\\
\hline	
%%%%%%%%%%%%%%%%%%%%%%%%%%%%%
	&	\textbf{SF begins in central cluster NGC 602}	&		\\
$\lesssim$ 5 Myr	&	Central PMS stars form, from optical CMD fitting	&	\citet{gouliermis_clustered_2012}	\\
2 -- 4 Myr	&	Central OB and PMS stars form, from optical CMD fitting	&	 \citet{carlson_panchromatic_2011}	\\
\hline
%%%%%%%%%%%%%%%%%%%%%%%%%%%%%
	&	\textbf{Subsequent SF in N90}	&		\\
2.5 Myr	&	Maximum SF rate reached, from optical PMS population	&	\citet{cignoni_star_2009}	\\
$\lesssim$ 2.5 Myr	&	PMS stars form in sub-clusters along rim, from optical CMD fitting	&	\citet{gouliermis_clustered_2012}	\\
2 Myr 
&	
Median apparent age of PMS stars in CO clumps, from optical CMD fitting	
&	This work	\\
1 -- 2 Myr & Ongoing intermediate-mass SF, from SED fitting of YSOs 
& This work \\
< 1 -- 2 Myr	&	YSOs form, from SED fitting of evolutionary phase	&	\citet{carlson_panchromatic_2011}	\\
\enddata	
\end{deluxetable*}
%%%%%%%%%%%%%%%%%%%%%%%%%%%%%%%%%%%%%%%%%%%%%%%%%%%%%%%%%%

We review the scenarios for the formation of N90 presented in previous studies in Table \ref{tab:timescale}, and add the results of this work.  A combination of stimulated and stochastic star formation in this region of the SMC Wing has been ongoing for at least 100 Myr, with a notable extended star formation event between 25--40 Myr ago \citep{ramachandran_testing_2019, fulmer_testing_2020}.  Between 7--8 Myr ago a collision occurred between $\sim 500$ pc components of \HI{} within the supergiant shell now identified as SGS 1 \citep{nigra_ngc_2008, fukui_formation_2020}.  Turbulence and compression stemming from this collision triggered the formation of NGC 602 3--5 Myr ago \citep{carlson_panchromatic_2011, gouliermis_clustered_2012} and subsequent creation of the \HII{} region N90.  The parsec-scale CO clumps to the north of N90 may retain signatures of the \HI{} collision in the form of inflated excitation temperatures and column densities, but determining whether a collision between shells \citep{nigra_ngc_2008} or clouds \citep{fukui_formation_2020} is more likely to be responsible is not yet possible.

Intermediate-mass YSOs have been forming along the \HII{} rim over the last 1--2 Myr  \citep{carlson_panchromatic_2011}, but it is unclear if this was triggered by the formation of the central cluster.  There is very little variation in the ages of PMS stars with distance from NGC 602, and although some isolated clusters of young ($\lesssim$2 Myr) PMS stars appear to exist along the rim, they coincide with CO emission sufficiently strong to cause age underestimates by $\gtrsim$1 Myr.  There is some evidence for YSOs disrupting their natal clumps on the parsec-scale, but on the scale of the entire region, there are very few significant correlations between clump properties and  distance from NGC 602 or Sk 183.  We conclude that a sequential star formation process, in which the creation of the central NGC 602 cluster did not directly cause the formation of the YSOs or PMS stars along the rim, is more likely to be present than a triggered scenario.

%%%%%%%%%%%%%%%%%%%%%%%%%%%%%%%%%%%%%%%%%%%%%%%%%%%%%%%%%%%%%%%%%%%%%%%%%%%%%%%%%%%%%%%%%%%%%
\subsection{Conclusions}

%%%%%%%%%%%%%%%%%%%%%%%%%%%%%%%%%%
We present results from ALMA observations of molecular gas in the low-metallicity star-forming region NGC 602/N90.  The main conclusions of this analysis are as follows:

\begin{enumerate}
    \item CO emission in N90 is confined to 110 sub-parsec-scale clumps arranged around the region's rim.  Only 26\% of clumps are traced by both $^{12}$CO and $^{13}$CO, with the remaining 74\% of clumps only being traced by $^{12}$CO with no strong corresponding $^{13}$CO emission (\S\ref{subsec:coldensity}).  We derive a  CO-to-H$_2$ conversion factor of $X_{\rm{CO, B}} = (3.4 \pm 0.2) \times 10^{20}$ cm$^{-2}$ (K km s$^{-1}$)$^{-1}$ $\simeq$ 1.7 $X_{\rm{CO, MW}}$ from the clumps that do possess strong emission in both $^{12}$CO and $^{13}$CO.  Applying this factor to all clumps yields a total CO-traced mass  of $3,310 \ \pm \ 250  \ M_\odot$ (\S\ref{subsec:xfactor}).

    \item We estimate a total molecular gas mass in N90 of $16,600 \pm 2,400 \ M_\odot$ through CO-dark gas mass correction (\S\ref{subsec:dark}).

    %%%%%%%%%%%%%%
    \item Clumps in N90 do not agree with expected trends in size-linewidth-surface density space, and have larger velocity dispersions and lower surface densities than predicted by relationships derived from Galactic clouds (\S\ref{subsec:sizeline}).  Additionally, CO-derived clump masses are significantly lower than virial masses, yielding high virial parameters (typical $\alpha_{\rm{vir}}=$ 4--11) and implying that clumps are either dispersing or confined by high levels of external pressure (\S\ref{subsec:virial}).  We use models of clumps with CO-dark gas to demonstrate that it is unlikely that CO-dark gas is responsible either of these effects.

    %%%%%%%%%%%%%%
    \item We refit \textit{Spitzer} YSO candidates identified by \citet{carlson_panchromatic_2011} and find by including new mid-to-far IR photometry that nearly all objects are less massive than previously estimated. Analysis of the present day accretion rate of the YSO candidate reveals that intermediate-mass star formation has likely been occurring throughout N90 in the last 1--2 Myr (\S\ref{S:yso}). 85\% of YSO candidates within the field observed by ALMA appear to be embedded within CO clumps.  We derive a recent ($\lesssim$1 Myr) SFR of $130 \pm 30 \ M_\odot$ Myr$^{-1}$, with a total YSO mass of $1250 \pm 160 \ M_\odot$ and CO-dark gas corrected SFE of $\epsilon \simeq 8 \pm 3 \%$ (\S\ref{s:SF}).
    
    %%%%%%%%%%%%%%
    \item We find no strong evidence that NGC 602 has directly triggered star formation along the rim of N90.  Spatial position relative to NGC 602 and the rim are poor predictor of clump properties (\S\ref{S:spatial}), as is association with YSOs or PMS stars (\S\ref{subsec:formation}), and there is no correlation between the age of PMS stars and radial distance from NGC 602.  Although some clusters of PMS stars along the rim appear marginally younger than PMS stars surrounding NGC 602 ($\sim$2 Myr vs 3 Myr), they are coincident with strong CO emission and thus the high extinction in these regions could cause their ages to appear younger than they truly are (\S\ref{S:pms}).
    
    %%%%%%%%%%%%%%
\end{enumerate}

Our analysis of the now-resolved sub-parsec-scale clumps in N90 has revealed the sequential star formation history of the region, and its evolution relative to the SMC Wing.  This more complete census of the total molecular gas mass in the region allows for an improved estimate of star formation efficiency on both by-clump and region scales.  After correction for CO-dark molecular gas content, we find that star formation in N90 is not more efficient than star formation in similarly massive solar-metallicity, higher-density environments.  

We consider N90 in the context of star formation in general in metal-poor environments.  Despite the low-metallicity and low-density environment of the SMC Wing, the properties of molecular clumps and SFE in N90 do not appear to dramatically differ from their Galactic counterparts.  If we extend this conclusion from the SMC to other nearby, small galaxies, it is likely that although star formation is initially sporadic in such environments, once regions have developed for a sufficient period of time their behavior does not depart significantly from the process of star formation in higher-metallicity, higher-density regions.

%%%%%%%%%%%%%%%%%%%%%%%%%%%%%%
\begin{acknowledgements}
%\vspace{4cm}

We thank Jay Gallagher for helpful discussions.  T.J.O. and R.I. were supported during this work by NSF award 2009624.  The material is partially based upon work supported by NASA under award number 80GSFC21M0002 (M.S.).  This paper makes use of the following ALMA data: ADS/JAO.ALMA\#2016.1.00360.S, ADS/JAO.ALMA\#2012.1.00683.S,  ADS/JAO.ALMA \#2015.1.01013.S,  ADS/JAO.ALMA\#2016.1.00193.S. ALMA is a partnership of ESO (representing its member states), NSF (USA) and NINS (Japan), together with NRC (Canada), MOST and ASIAA (Taiwan), and KASI (Republic of Korea), in cooperation with the Republic of Chile.  The Joint ALMA Observatory is operated by ESO, AUI/NRAO and NAOJ.  The National Radio Astronomy Observatory is a facility of the National Science Foundation operated under cooperative agreement by Associated Universities, Inc.  

\end{acknowledgements}

%%%%%%%%%%%%%%%%%%%%%%%%%%%%%%
\software{APLpy \citep{aplpy2012,aplpy2019};  
astrodendro \citep{Robitaille2019_astrodendro};
 Astropy \citep{astropy_2013,astropy_2018};
 CASA \citep{McMullin2007_casa};
 DOLPHOT \citep{dolphot_dolphin2016}; 
Matplotlib \citep{matplotlib_Hunter2007}; 
Numpy \citep{harris2020_numpy}; 
OpenCV \citep{opencv_library};
Pandas \citep{pandas_mckinney-proc-scipy-2010,pandas_reback2020}; quickclump \citep{Sidorin2017_quickclump};
Seaborn \citep{seaborn_Waskom2021};
scikit-learn \citep{scikit-learn};
Scipy \citep{2020SciPy-NMeth};
spectral$\_$cube \citep{spectral_cube_ginsburg2019};
statsmodels \citep{statsmodels_seabold2010};
TA-DA \citep{da_rio_ta-da_2012}}

\facilities{ALMA, HST, Spitzer, Herschel}

%%%%%%%%%%%%%%%%%%%%%%%%%%%%%%%%%%%%%%%%%%%%%%%%%%%%%%%%%%%%
\appendix
\restartappendixnumbering
%%%%%%%%%%%%%%%%%%%%%%%%%%%%%%%%%%%%%%%%%%%%%%%%%%%%%%%%%%%%

%%%%%%%%%%%%%%%%%%%%%%%%%%%%
\section{Clump and YSO Properties}\label{ap:tabs}

We present tables of the properties of CO clumps (Table \ref{tab:clumps}) and YSO candidates (Table \ref{tab:yso}) in N90.

\begin{longrotatetable}
\begin{deluxetable}{ccccccccccccccccc}
\tabletypesize{\scriptsize}
\tablecaption{Clump Properties \label{tab:clumps}} 
\tablehead{\colhead{ID} &  \colhead{RA} & \colhead{Dec} & \colhead{$^{12}$CO$_{\rm{pk}}$} & 
  \colhead{$^{13}$CO$_{\rm{pk}}$} & 
  \colhead{$W_{12}$} & 
  \colhead{$R$} & 
  \colhead{v$_{\rm{LSRK, 12}}$} &
  \colhead{$\sigma_{v,12}$} & 
  \colhead{M$_{\rm{LTE}}$} &
  \colhead{M$_{X_{CO,B}}$} &
   \colhead{$\Sigma$} & 
   \colhead{$\rho_c$} & 
  \colhead{M$_{\rm{vir}}$} &
   \colhead{$\alpha_{\rm{vir}}$} &  
  \colhead{Assoc. YSO} & 
  \colhead{Group} \\[-0.1em]   
 \colhead{}   & \colhead{($\degr$)}  & \colhead{($\degr$)} & 
 \colhead{(K)} & \colhead{(K)} & 
 \colhead{(K km s$^{-1}$)} & 
 \colhead{(pc)}  &
 \colhead{(km s$^{-1}$)} &
  \colhead{(km s$^{-1}$)} &
\colhead{$(M_{\odot})$} &   
\colhead{$(M_{\odot})$} & 
\colhead{$(M_{\odot}$ pc$^{-2})$} & 
\colhead{$(M_{\odot}$ pc$^{-3})$} & 
\colhead{$(M_{\odot})$} &
\colhead{} &
  \colhead{}  & \colhead{}  }
\startdata
       1 & 22.3982 & -73.5501 &   22.55 &    6.26 &  9355.61 & 0.55 &    171.88 &    0.90 &  245.1 & 188.9 &  201.7 & 1008.6 &   464.7 &         2.5 &            &       \\
       2 & 22.3986 & -73.5581 &   21.85 &    3.79 &  5146.64 & 0.62 &    173.74 &    0.56 &  102.6 & 103.9 &   86.8 &  434.2 &   203.1 &         2.0 &       Y270 &     r \\
       3 & 22.3990 & -73.5506 &   20.20 &    7.11 & 13901.61 & 0.61 &    173.55 &    0.93 &  286.1 & 280.7 &  243.4 & 1217.1 &   544.6 &         1.9 &       Y327 &       \\
       4 & 22.3984 & -73.5610 &   20.01 &    4.96 &  7215.35 & 0.50 &    174.96 &    1.07 &  163.6 & 145.7 &  183.7 &  918.7 &   603.0 &         4.1 &       Y251 &     r \\
       5 & 22.4135 & -73.5553 &   18.15 &    4.97 &  5699.55 & 0.58 &    171.30 &    0.61 &  143.6 & 115.1 &  109.1 &  545.3 &   227.7 &         2.0 &       Y312 &       \\
       6 & 22.4002 & -73.5647 &   16.54 &    4.24 &  4298.67 & 0.49 &    168.72 &    0.71 &   83.5 &  86.8 &  116.9 &  584.3 &   257.2 &         3.0 &       Y223 &     r \\
       7 & 22.4208 & -73.5583 &   16.41 &    2.66 &  3721.89 & 0.64 &    171.09 &    0.38 &   44.7 &  75.2 &   57.6 &  287.9 &    96.8 &         1.3 &            &       \\
       8 & 22.4314 & -73.5622 &   16.23 &    3.99 &  4976.71 & 0.68 &    166.36 &    0.48 &   61.9 & 100.5 &   68.4 &  341.8 &   166.7 &         1.7 &       Y287 &       \\
       9 & 22.4001 & -73.5644 &   16.22 &    3.69 &  3596.83 & 0.51 &    167.30 &    0.69 &   65.5 &  72.6 &   87.5 &  437.4 &   258.7 &         3.6 &            &     r \\
      10 & 22.4369 & -73.5573 &   15.37 &    4.71 &  5210.15 & 0.61 &    166.32 &    0.65 &  123.3 & 105.2 &   88.9 &  444.5 &   268.0 &         2.5 &       Y326 &       \\
      11 & 22.4234 & -73.5622 &   15.13 &    3.71 &  7338.28 & 0.76 &    166.79 &    0.56 &  153.5 & 148.2 &   81.8 &  409.1 &   251.7 &         1.7 &            &       \\
      12 & 22.4053 & -73.5649 &   14.72 &    1.86 &  7512.51 & 0.77 &    169.59 &    0.64 &   91.6 & 151.7 &   82.5 &  412.6 &   324.2 &         2.1 & Y227, Y240 &     r \\
      13 & 22.3320 & -73.5535 &   14.42 &    2.79 &  7856.07 & 0.77 &    161.21 &    0.88 &  142.5 & 158.6 &   85.3 &  426.6 &   619.5 &         3.9 &       Y217 &     r \\
      14 & 22.4049 & -73.5643 &   14.19 &    2.26 &  3586.16 & 0.47 &    168.89 &    0.75 &   56.4 &  72.4 &  102.3 &  511.6 &   281.7 &         3.9 &            &     r \\
      15 & 22.4336 & -73.5621 &   13.88 &    2.41 &  1850.34 & 0.54 &    166.14 &    0.38 &   26.9 &  37.4 &   40.1 &  200.6 &    82.0 &         2.2 &            &       \\
      16 & 22.3601 & -73.5416 &   13.51 &    1.76 &  2758.22 & 0.51 &    167.72 &    0.59 &   64.0 &  55.7 &   67.9 &  339.5 &   186.1 &         3.3 &       A340 &       \\
      17 & 22.4277 & -73.5617 &   13.29 &    2.95 &  3921.42 & 0.52 &    166.55 &    0.59 &   82.5 &  79.2 &   93.6 &  467.9 &   190.3 &         2.4 &            &       \\
      18 & 22.3456 & -73.5475 &   13.23 &    3.63 &  4710.30 & 0.53 &    162.64 &    0.67 &   62.6 &  95.1 &  106.7 &  533.3 &   250.4 &         2.6 &       Y285 &     r \\
      19 & 22.3787 & -73.5684 &   13.07 &    2.92 &  3776.69 & 0.58 &    164.70 &    0.68 &  101.9 &  76.3 &   72.2 &  360.8 &   278.6 &         3.7 &       Y170 &     r \\
      20 & 22.4189 & -73.5547 &   12.43 &    1.48 &  2580.00 & 0.53 &    170.18 &    0.53 &      - &  52.1 &   59.7 &  298.4 &   157.5 &         3.0 &            &       \\
      21 & 22.3354 & -73.5518 &   10.94 &    3.84 &  2631.43 & 0.42 &    162.40 &    0.92 &   78.2 &  53.1 &   95.3 &  476.3 &   369.6 &         7.0 &       Y237 &     r \\
      22 & 22.3991 & -73.5614 &   10.58 &    1.83 &  1345.89 & 0.37 &    177.47 &    0.54 &   34.8 &  27.2 &   64.3 &  321.4 &   112.9 &         4.2 &            &     r \\
      23 & 22.4154 & -73.5547 &   10.57 &    2.91 &  1717.81 & 0.44 &    171.33 &    0.62 &   26.4 &  34.7 &   56.8 &  283.8 &   178.0 &         5.1 &            &       \\
      24 & 22.3575 & -73.5798 &   10.07 &    2.68 &  1896.67 & 0.57 &    166.28 &    0.34 &   40.2 &  38.3 &   37.4 &  187.1 &    67.7 &         1.8 &       Y090 &       \\
      25 & 22.4136 & -73.5715 &   10.00 &    1.19 &  1000.70 & 0.38 &    167.03 &    0.47 &      - &  20.2 &   44.0 &  220.0 &    88.9 &         4.4 &            &       \\
      26 & 22.4104 & -73.5646 &    9.92 &    1.35 &  1208.22 & 0.42 &    169.67 &    0.45 &      - &  24.4 &   44.4 &  222.0 &    89.6 &         3.7 &            &     r \\
      27 & 22.4007 & -73.5723 &    9.88 &    1.70 &   774.79 & 0.40 &    168.02 &    0.38 &   12.7 &  15.6 &   31.0 &  155.2 &    59.7 &         3.8 &            &       \\
      28 & 22.3326 & -73.5521 &    9.77 &    1.85 &  2631.80 & 0.52 &    161.14 &    0.73 &   39.0 &  53.1 &   62.7 &  313.4 &   287.3 &         5.4 &            &     r \\
      29 & 22.4189 & -73.5569 &    9.74 &    1.19 &   740.74 & 0.40 &    171.20 &    0.33 &      - &  15.0 &   29.6 &  148.2 &    45.2 &         3.0 &            &       \\
      30 & 22.4168 & -73.5546 &    9.73 &    1.78 &  1530.42 & 0.43 &    170.20 &    0.56 &   23.5 &  30.9 &   54.2 &  270.8 &   140.1 &         4.5 &            &       \\
      31 & 22.3591 & -73.5762 &    9.02 &    1.35 &  1653.76 & 0.40 &    165.93 &    0.69 &      - &  33.4 &   66.9 &  334.5 &   198.7 &         6.0 &            &       \\
      32 & 22.3323 & -73.5563 &    9.02 &    1.25 &  1757.42 & 0.48 &    162.98 &    0.67 &      - &  35.5 &   48.5 &  242.7 &   229.8 &         6.5 &       Y196 &     r \\
      33 & 22.3817 & -73.5694 &    8.74 &    2.81 &   892.46 & 0.36 &    171.44 &    0.58 &   28.9 &  18.0 &   43.2 &  216.2 &   130.1 &         7.2 &       Y162 &     r \\
      34 & 22.4150 & -73.5647 &    8.00 &    1.14 &   686.99 & 0.40 &    169.48 &    0.36 &      - &  13.9 &   27.7 &  138.6 &    53.0 &         3.8 &            &     r \\
      35 & 22.3977 & -73.5684 &    7.91 &    1.51 &   732.43 & 0.38 &    170.33 &    0.45 &      - &  14.8 &   32.3 &  161.7 &    81.3 &         5.5 &            &     r \\
      36 & 22.3996 & -73.5734 &    7.90 &    1.27 &   943.82 & 0.44 &    166.46 &    0.43 &      - &  19.1 &   31.7 &  158.7 &    84.3 &         4.4 &            &       \\
      37 & 22.3955 & -73.5677 &    7.73 &    1.16 &   805.49 & 0.38 &    171.55 &    0.49 &      - &  16.3 &   35.6 &  177.8 &    94.8 &         5.8 &            &     r \\
      38 & 22.3849 & -73.5691 &    7.55 &    3.35 &  1028.77 & 0.37 &    170.68 &    0.68 &   34.3 &  20.8 &   47.5 &  237.4 &   182.3 &         8.8 &       Y171 &     r \\
      39 & 22.3970 & -73.5643 &    7.37 &    1.38 &   842.36 & 0.42 &    166.68 &    0.53 &      - &  17.0 &   30.5 &  152.7 &   123.8 &         7.3 &            &     r \\
      40 & 22.3919 & -73.5699 &    7.30 &    1.04 &   905.78 & 0.43 &    170.20 &    0.50 &      - &  18.3 &   31.2 &  156.0 &   111.6 &         6.1 &       Y174 &     r \\
      41 & 22.4052 & -73.5700 &    7.26 &    1.38 &   697.39 & 0.41 &    170.67 &    0.45 &      - &  14.1 &   26.1 &  130.7 &    87.5 &         6.2 &            &       \\
      42 & 22.3303 & -73.5558 &    7.00 &    1.55 &  1461.28 & 0.51 &    162.82 &    0.69 &      - &  29.5 &   35.6 &  177.8 &   259.0 &         8.8 &       Y197 &     r \\
      43 & 22.3272 & -73.5565 &    6.98 &    1.23 &   934.29 & 0.42 &    163.04 &    0.57 &      - &  18.9 &   33.3 &  166.5 &   144.1 &         7.6 &       Y179 &     r \\
      44 & 22.4006 & -73.5748 &    6.92 &    1.39 &  1363.03 & 0.62 &    166.89 &    0.67 &      - &  27.5 &   22.5 &  112.7 &   289.2 &        10.5 &            &       \\
      45 & 22.3971 & -73.5735 &    6.89 &    1.15 &  1016.27 & 0.43 &    165.09 &    0.65 &      - &  20.5 &   35.6 &  177.8 &   190.1 &         9.3 &       Y148 &       \\
      46 & 22.3926 & -73.5594 &    6.79 &    1.31 &   691.45 & 0.40 &    177.65 &    0.38 &      - &  14.0 &   27.4 &  137.2 &    60.9 &         4.4 &            &     r \\
      47 & 22.4063 & -73.5663 &    6.74 &    1.38 &  1452.12 & 0.48 &    170.65 &    0.57 &      - &  29.3 &   40.1 &  200.6 &   164.4 &         5.6 &            &     r \\
      48 & 22.3403 & -73.5440 &    6.72 &    1.55 &  1711.77 & 0.53 &    162.08 &    0.56 &      - &  34.6 &   39.6 &  198.2 &   175.5 &         5.1 &            &     r \\
      49 & 22.3350 & -73.5548 &    6.70 &    1.53 &   832.18 & 0.37 &    164.05 &    0.61 &      - &  16.8 &   38.8 &  193.9 &   146.9 &         8.7 &       Y206 &     r \\
      50 & 22.3423 & -73.5468 &    6.55 &    1.28 &   952.40 & 0.43 &    160.74 &    0.61 &      - &  19.2 &   32.6 &  163.1 &   169.7 &         8.8 &            &     r \\
      51 & 22.3896 & -73.5740 &    6.42 &    1.12 &   614.31 & 0.41 &    162.82 &    0.44 &      - &  12.4 &   23.7 &  118.3 &    81.4 &         6.6 &            &       \\
      52 & 22.4097 & -73.5641 &    6.11 &    1.19 &   460.36 & 0.35 &    168.86 &    0.46 &      - &   9.3 &   23.7 &  118.3 &    77.3 &         8.3 &            &     r \\
      53 & 22.3946 & -73.5736 &    5.67 &    0.90 &   509.58 & 0.43 &    164.36 &    0.37 &      - &  10.3 &   17.6 &   88.1 &    61.1 &         5.9 &            &       \\
      54 & 22.4073 & -73.5660 &    5.67 &    1.26 &  1443.38 & 0.42 &    168.87 &    0.73 &      - &  29.1 &   52.8 &  264.2 &   232.7 &         8.0 &            &     r \\
      55 & 22.4035 & -73.5640 &    5.67 &    1.81 &   363.58 & 0.33 &    175.09 &    0.46 &   16.3 &   7.3 &   21.2 &  105.8 &    73.6 &        10.0 &            &     r \\
      56 & 22.4025 & -73.5727 &    5.65 &    1.20 &   559.86 & 0.40 &    168.54 &    0.49 &      - &  11.3 &   22.4 &  111.8 &    99.0 &         8.8 &            &       \\
      57 & 22.3349 & -73.5460 &    5.60 &    1.08 &   528.11 & 0.39 &    161.49 &    0.45 &      - &  10.7 &   21.8 &  108.9 &    84.7 &         7.9 &            &     r \\
      58 & 22.4054 & -73.5650 &    5.52 &    1.19 &   669.50 & 0.40 &    172.79 &    0.56 &      - &  13.5 &   26.7 &  133.5 &   132.5 &         9.8 &            &     r \\
      59 & 22.3391 & -73.5456 &    5.48 &    1.03 &   353.19 & 0.34 &    160.94 &    0.37 &      - &   7.1 &   19.3 &   96.7 &    48.8 &         6.8 &            &     r \\
      60 & 22.4206 & -73.5718 &    5.39 &    0.85 &   297.48 & 0.39 &    166.52 &    0.28 &      - &   6.0 &   12.7 &   63.6 &    32.7 &         5.4 &            &       \\
      61 & 22.3861 & -73.5759 &    5.37 &    0.96 &   253.89 & 0.35 &    169.09 &    0.33 &      - &   5.1 &   13.5 &   67.7 &    39.6 &         7.7 &            &       \\
      62 & 22.4055 & -73.5647 &    5.19 &    1.03 &   330.24 & 0.37 &    171.85 &    0.37 &      - &   6.7 &   15.8 &   79.1 &    53.0 &         7.9 &            &     r \\
      63 & 22.4037 & -73.5653 &    5.19 &    1.06 &   943.70 & 0.43 &    166.95 &    0.64 &      - &  19.1 &   32.7 &  163.4 &   186.7 &         9.8 &            &     r \\
      64 & 22.3686 & -73.5756 &    5.12 &    0.91 &   325.28 & 0.37 &    163.52 &    0.37 &      - &   6.6 &   15.5 &   77.6 &    51.3 &         7.8 &            &       \\
      65 & 22.4047 & -73.5600 &    4.97 &    1.30 &   329.91 & 0.34 &    168.44 &    0.46 &      - &   6.7 &   18.0 &   89.8 &    76.3 &        11.5 &       Y264 &     r \\
      66 & 22.3406 & -73.5466 &    4.77 &    1.08 &   464.89 & 0.45 &    161.06 &    0.43 &      - &   9.4 &   14.8 &   73.9 &    85.9 &         9.1 &            &     r \\
      67 & 22.3292 & -73.5644 &    4.62 &    1.28 &   548.05 & 0.42 &    162.92 &    0.49 &      - &  11.1 &   19.5 &   97.6 &   107.8 &         9.7 &            &     r \\
      68 & 22.3265 & -73.5572 &    4.41 &    1.39 &   740.07 & 0.42 &    162.90 &    0.67 &      - &  14.9 &   27.1 &  135.5 &   195.5 &        13.1 &            &     r \\
      69 & 22.3185 & -73.5669 &    4.39 &    1.15 &   293.45 & 0.37 &    165.99 &    0.39 &      - &   5.9 &   13.6 &   68.1 &    59.8 &        10.1 &            &     r \\
      70 & 22.4111 & -73.5636 &    4.31 &    1.36 &   217.83 & 0.32 &    165.19 &    0.40 &      - &   4.4 &   13.7 &   68.7 &    54.7 &        12.4 &            &     r \\
      71 & 22.4083 & -73.5658 &    4.26 &    0.96 &   513.34 & 0.41 &    167.73 &    0.40 &      - &  10.4 &   20.0 &   99.8 &    69.7 &         6.7 &            &     r \\
      72 & 22.3363 & -73.5683 &    4.22 &    1.15 &   464.06 & 0.43 &    162.36 &    0.53 &      - &   9.4 &   16.4 &   82.1 &   126.3 &        13.5 &            &       \\
      73 & 22.4065 & -73.5619 &    4.20 &    0.81 &   194.54 & 0.31 &    179.52 &    0.30 &      - &   3.9 &   12.7 &   63.4 &    30.5 &         7.8 &       Y255 &     r \\
      74 & 22.4224 & -73.5715 &    4.08 &    1.02 &   209.90 & 0.39 &    166.81 &    0.24 &      - &   4.2 &    9.0 &   45.1 &    22.9 &         5.4 &            &       \\
      75 & 22.3974 & -73.5773 &    4.04 &    1.05 &   433.77 & 0.49 &    167.69 &    0.28 &      - &   8.8 &   11.8 &   59.2 &    38.8 &         4.4 &            &       \\
      76 & 22.3687 & -73.5699 &    4.03 &    0.97 &   420.55 & 0.43 &    172.38 &    0.47 &      - &   8.5 &   15.0 &   74.8 &    98.0 &        11.5 &       Y142 &     r \\
      77 & 22.3818 & -73.5721 &    3.99 &    1.10 &   215.29 & 0.35 &    171.62 &    0.36 &      - &   4.3 &   11.5 &   57.5 &    47.8 &        11.0 &            &       \\
      78 & 22.3841 & -73.5757 &    3.98 &    1.65 &   139.30 & 0.30 &    164.30 &    0.29 &      - &   2.8 &    9.6 &   48.2 &    26.7 &         9.5 &            &       \\
      79 & 22.3734 & -73.5751 &    3.93 &    1.35 &   179.92 & 0.32 &    170.87 &    0.34 &      - &   3.6 &   11.4 &   56.9 &    38.1 &        10.5 &       Y118 &       \\
      80 & 22.3435 & -73.5457 &    3.82 &    1.29 &   196.62 & 0.33 &    160.71 &    0.34 &      - &   4.0 &   11.4 &   57.0 &    41.3 &        10.4 &            &     r \\
      81 & 22.3399 & -73.5450 &    3.81 &    1.05 &   718.68 & 0.41 &    162.25 &    0.73 &      - &  14.5 &   27.5 &  137.3 &   228.4 &        15.7 &            &     r \\
      82 & 22.3666 & -73.5751 &    3.76 &    1.04 &   180.02 & 0.36 &    165.65 &    0.32 &      - &   3.6 &    8.9 &   44.4 &    38.8 &        10.7 &            &       \\
      83 & 22.4257 & -73.5635 &    3.74 &    1.10 &   151.55 & 0.33 &    169.65 &    0.38 &      - &   3.1 &    8.7 &   43.7 &    50.4 &        16.5 &            &       \\
      84 & 22.3195 & -73.5611 &    3.66 &    0.86 &   206.17 & 0.36 &    162.19 &    0.40 &      - &   4.2 &   10.1 &   50.3 &    59.8 &        14.4 &       Y143 &     r \\
      85 & 22.4143 & -73.5649 &    3.63 &    0.89 &   123.62 & 0.31 &    167.28 &    0.33 &      - &   2.5 &    8.2 &   41.1 &    34.7 &        13.9 &            &     r \\
      86 & 22.3300 & -73.5549 &    3.56 &    1.24 &   322.38 & 0.40 &    162.55 &    0.42 &      - &   6.5 &   12.9 &   64.7 &    72.9 &        11.2 &            &     r \\
      87 & 22.3910 & -73.5600 &    3.56 &    1.12 &   162.66 & 0.33 &    177.21 &    0.38 &      - &   3.3 &    9.8 &   48.8 &    50.5 &        15.4 &            &     r \\
      88 & 22.4038 & -73.5721 &    3.54 &    0.75 &   164.01 & 0.31 &    168.88 &    0.33 &      - &   3.3 &   10.8 &   54.1 &    35.5 &        10.7 &            &       \\
      89 & 22.3265 & -73.5649 &    3.47 &    1.03 &   184.61 & 0.33 &    162.45 &    0.41 &      - &   3.7 &   11.1 &   55.3 &    57.8 &        15.5 &            &     r \\
      90 & 22.4005 & -73.5745 &    3.20 &    0.91 &   174.52 & 0.37 &    164.11 &    0.31 &      - &   3.5 &    8.0 &   40.2 &    38.4 &        10.9 &            &       \\
      91 & 22.3953 & -73.5727 &    3.16 &    0.72 &   120.61 & 0.26 &    165.04 &    0.35 &      - &   2.4 &   11.1 &   55.6 &    33.1 &        13.6 &            &       \\
      92 & 22.3961 & -73.5563 &    3.11 &    0.73 &    89.55 & 0.31 &    177.72 &    0.28 &      - &   1.8 &    5.8 &   29.2 &    26.0 &        14.4 &            &     r \\
      93 & 22.3255 & -73.5644 &    3.07 &    1.26 &   308.87 & 0.38 &    163.82 &    0.50 &      - &   6.2 &   13.8 &   68.8 &    99.6 &        16.0 &            &     r \\
      94 & 22.3362 & -73.5476 &    2.95 &    1.02 &    81.51 & 0.30 &    162.29 &    0.28 &      - &   1.6 &    5.9 &   29.3 &    25.2 &        15.3 &            &     r \\
      95 & 22.3087 & -73.5570 &    2.92 &    0.98 &    72.28 & 0.30 &    162.47 &    0.23 &      - &   1.5 &    5.0 &   25.1 &    16.5 &        11.3 &            &       \\
      96 & 22.3258 & -73.5522 &    2.91 &    1.32 &    80.05 & 0.31 &    161.55 &    0.24 &      - &   1.6 &    5.4 &   27.1 &    19.3 &        11.9 &            &     r \\
      97 & 22.3906 & -73.5716 &    2.90 &    0.87 &    83.71 & 0.31 &    163.52 &    0.25 &      - &   1.7 &    5.6 &   27.9 &    20.5 &        12.1 &            &       \\
      98 & 22.3394 & -73.5430 &    2.81 &    0.52 &   123.97 & 0.35 &    160.78 &    0.30 &      - &   2.5 &    6.5 &   32.4 &    33.1 &        13.2 &            &     r \\
      99 & 22.4028 & -73.5731 &    2.76 &    0.85 &    82.34 & 0.28 &    167.84 &    0.25 &      - &   1.7 &    6.5 &   32.6 &    19.1 &        11.5 &            &       \\
     100 & 22.3662 & -73.5755 &    2.74 &    1.04 &    84.38 & 0.26 &    164.22 &    0.27 &      - &   1.7 &    7.7 &   38.7 &    20.1 &        11.8 &            &       \\
     101 & 22.3162 & -73.5620 &    2.69 &    0.83 &   170.67 & 0.37 &    161.04 &    0.35 &      - &   3.4 &    8.2 &   41.2 &    46.7 &        13.5 &            &     r \\
     102 & 22.4179 & -73.5538 &    2.68 &    0.82 &    94.77 & 0.28 &    171.81 &    0.32 &      - &   1.9 &    7.8 &   39.0 &    29.4 &        15.3 &            &       \\
     103 & 22.4082 & -73.5589 &    2.65 &    1.22 &    98.20 & 0.28 &    168.46 &    0.38 &      - &   2.0 &    7.9 &   39.7 &    43.2 &        21.8 &            &     r \\
     104 & 22.3223 & -73.5644 &    2.65 &    0.57 &   103.14 & 0.30 &    164.09 &    0.35 &      - &   2.1 &    7.2 &   35.8 &    39.2 &        18.8 &            &     r \\
     105 & 22.3216 & -73.5625 &    2.58 &    0.98 &   396.29 & 0.47 &    162.03 &    0.51 &      - &   8.0 &   11.7 &   58.3 &   127.9 &        16.0 &            &     r \\
     106 & 22.3683 & -73.5777 &    2.56 &    1.13 &   121.61 & 0.34 &    170.15 &    0.35 &      - &   2.5 &    6.9 &   34.5 &    44.2 &        18.0 &            &       \\
     107 & 22.3374 & -73.5462 &    2.56 &    1.27 &   204.92 & 0.37 &    159.52 &    0.64 &      - &   4.1 &    9.4 &   47.2 &   161.5 &        39.0 &            &     r \\
     108 & 22.3201 & -73.5615 &    2.50 &    1.04 &    93.63 & 0.35 &    160.73 &    0.28 &      - &   1.9 &    4.9 &   24.6 &    28.6 &        15.1 &            &     r \\
     109 & 22.3906 & -73.5732 &    2.36 &    0.81 &    59.62 & 0.28 &    164.27 &    0.24 &      - &   1.2 &    4.8 &   24.1 &    17.6 &        14.6 &            &       \\
     110 & 22.3332 & -73.5466 &    2.25 &    0.69 &    69.29 & 0.37 &    160.04 &    0.26 &      - &   1.4 &    3.3 &   16.6 &    25.4 &        18.1 &            &     r   
\enddata
\tablecomments{Clumps without LTE mass estimates have $^{13}$CO S/N < 3.  $\rho_c$ is the volume density at 0.1 pc.  ``Assoc. YSO`` gives the first four characters of the identifiers assigned by \citet{carlson_panchromatic_2011} and listed in Table \ref{tab:yso} of any YSOs which fall within the projected 2D boundaries of the clump.  A denotation of 'r' in Group indicates the clump is on the N90 rim.  Error estimates for most quantities are available in a complete, machine-readable version of this table.  }

\end{deluxetable}
\end{longrotatetable}

%%%%%%%%%%%%%%%%%%%%%%%%%%%%%%%%%%%%%%%%%%%%

\startlongtable
\begin{deluxetable*}{cccccc}
\tabletypesize{\scriptsize}
\tablecaption{YSO Properties \label{tab:yso}} 
\tablehead{
  \colhead{ID} & \colhead{Name} & \colhead{RA} & \colhead{Dec} & \colhead{M} &  \colhead{$\dot{M}$} \\
  \vspace{-0.4cm} \\
  \colhead{} & \colhead{}  & \colhead{($\degr$)}  & \colhead{($\degr$)} & \colhead{(M$_\odot$)}  & \colhead{($10^{-6}$ M$_\odot$ yr$^{-1}$)} \\  \vspace{-0.4cm} }
\startdata
Y090 & J012925.97-733446.8 &  22.3582 & -73.5797 &  3.71 &   35.58 \\
Y096 & J012906.41-733348.6 &  22.2767 & -73.5635 &  7.78 &    3.50 \\
Y118 & J012929.62-733430.1 &  22.3734 & -73.5750 &  2.00 &    0.36 \\ 
Y142 & J012928.56-733411.9 &  22.3690 & -73.5700 &  3.54 &    1.93 \\
Y143 & J012916.77-733340.7 &  22.3199 & -73.5613 &  2.02 &    0.19 \\
Y148 & J012935.11-733423.9 &  22.3963 & -73.5733 &  3.64 &    7.48 \\ 
Y149 & J012856.16-733242.4 &  22.2340 & -73.5451 &  2.93 &    0.33 \\ 
Y162 & J012931.68-733409.2 &  22.3820 & -73.5692 &  2.91 &   91.79 \\
Y163 & J012859.33-733244.8 &  22.2472 & -73.5458 &  2.89 &    0.43 \\ 
Y170 & J012930.90-733405.6 &  22.3788 & -73.5682 &  3.33 &   11.20 \\ 
Y171 & J012932.39-733408.4 &  22.3850 & -73.5690 &  3.28 &    6.61 \\
Y174 & J012934.05-733411.7 &  22.3919 & -73.5699 &  3.58 &    0.94 \\
Y179 & J012918.44-733324.9 &  22.3269 & -73.5569 &  2.24 &    4.22 \\
Y196 & J012919.88-733322.5 &  22.3328 & -73.5563 &  3.76 &    0.75 \\
Y197 & J012918.95-733319.4 &  22.3290 & -73.5554 &  3.57 &    1.18 \\
Y198 & J012936.38-733403.6 &  22.4016 & -73.5677 &  5.94 &    6.05 \\ 
Y206 & J012920.48-733316.8 &  22.3353 & -73.5547 &  2.88 &    1.75 \\
Y217 & J012919.87-733312.5 &  22.3328 & -73.5535 &  3.88 &    1.40 \\
Y223 & J012935.89-733351.7 &  22.3996 & -73.5644 &  3.93 &   34.07 \\ 
Y227 & J012937.37-733352.4 &  22.4057 & -73.5646 &  6.99 &    0.38 \\ 
Y237 & J012920.64-733306.3 &  22.3360 & -73.5518 &  4.14 &    3.56 \\
Y240 & J012937.99-733352.8 &  22.4083 & -73.5647 &  6.17 &    4.24 \\ 
Y251 & J012935.64-733339.5 &  22.3985 & -73.5610 &  3.50 &   16.98 \\ 
Y255 & J012937.60-733342.9 &  22.4067 & -73.5619 &  3.23 &    4.28 \\ 
Y264 & J012936.99-733336.1 &  22.4041 & -73.5600 &  3.20 &    0.29 \\
A270 & J012935.51-733330.3 &  22.3980 & -73.5584 &  6.48 &   36.11 \\
Y270i & J012935.51-733330.3 &  22.3980 & -73.5584 &  6.16 &  251.03 \\ 
Y271 & J012933.43-733323.3 &  22.3893 & -73.5565 &  2.32 &    0.68 \\ 
Y283 & J012930.20-733310.5 &  22.3759 & -73.5529 &  2.23 &    0.76 \\  
Y285 & J012923.06-733251.5 &  22.3461 & -73.5476 &  2.92 &    0.07 \\  
Y287 & J012943.36-733343.1 &  22.4307 & -73.5620 &  3.19 &    0.24 \\ 
Y288 & J012942.42-733341.6 &  22.1768 & -73.0120 &  3.58 &   32.66 \\ 
Y290 & J012937.08-733325.4 &  22.4045 & -73.5570 &  3.50 &    0.08 \\  
Y312 & J012939.17-733318.9 &  22.4132 & -73.5553 &  3.31 &    0.39 \\  
Y326 & J012944.75-733325.2 &  22.4365 & -73.5570 &  2.79 &    2.51 \\ 
Y327 & J012935.69-733302.1 &  22.3987 & -73.5506 &  6.45 &    2.27 \\  
Y340 & J012926.56-733230.3 &  22.3607 & -73.5418 &  2.33 &    0.17 \\  
A340 & J012926.56-733230.3 &  22.3607 & -73.5418 &  2.77 &    1.89 \\ 
Y358 & J012935.17-733242.5 &  22.3966 & -73.5452 &  2.19 &    2.70 \\ 
Y396 & J012924.20-733152.8 &  22.3508 & -73.5313 &  2.37 &    0.21 \\  
Y493 & J012915.95-733017.6 &  22.3165 & -73.5049 &  1.60 &    0.03 \\ 
Y700 & J013006.79-733258.9 &  22.5286 & -73.5497 &  2.56 &    5.35 \\ 
K340 & J012926.56-733230.3 &  22.3607 & -73.5418 &  3.53 &    0.03 \\ 
U364 & J012908.96-733129.5 &  22.2873 & -73.5249 &  2.41 &    2.81 \\ 
U703 & J012911.03-733039.6 &  22.2961 & -73.5110 &  1.21 &    0.02 
 \\
\enddata
\tablecomments{IDs are as given in \citet{carlson_panchromatic_2011}.}
\end{deluxetable*}

%%%%%%%%%%%%%%%%%%%%%%%%%%%%%%
\bibliographystyle{yahapj}
\bibliography{refs.bib}

%%%%%%%%%%
\end{document}